\documentclass[twocolumn,superscriptaddress,aps]{revtex4-1}
\usepackage{mathrsfs}
\usepackage{amsmath}
\usepackage{mathtools}
\usepackage{graphicx}
\usepackage{amstext}
\usepackage{amsfonts}
\usepackage{amssymb}
\usepackage{bm}
\usepackage[usenames,dvipsnames]{color}
\usepackage{hyperref}
\usepackage{calc}

\newcommand{\+}{\dagger}
\newcommand{\ra}{\rightarrow}
\newcommand{\f}{\frac}
\newcommand{\8}{\infty}
\newcommand{\ket}{\rangle}
\newcommand{\bra}{\langle}
\newcommand{\pd}{\partial}
\newcommand{\X}{\otimes}

\newcommand{\w}{\omega}
\newcommand{\tr}{{\rm tr}}
\newcommand{\Tr}{{\rm Tr}}
\newcommand{\s}{{_{\!S}}}
\newcommand{\e}{{_{\!E}}}
\newcommand{\matx}[1]{\begin{pmatrix} #1 \end{pmatrix}}
\newcommand{\mat}[1]{\begin{pmatrix} #1 \end{pmatrix}}

\begin{document}

\title{Nonperturbative renormalization of  quantum thermodynamics \\ from weak to strong couplings}
\author{Wei-Ming Huang}
\affiliation{Department of Physics and Center for Quantum Information Science,
		National Cheng Kung University, Tainan 70101, Taiwan}
\author{Wei-Min Zhang}
\email{Correspondence author: wzhang@mail.ncku.edu.tw}
\affiliation{Department of Physics and Center for Quantum Information Science,
		National Cheng Kung University, Tainan 70101, Taiwan}
\affiliation{Physics Division, National Center for Theoretical Sciences, Taipei 10617, Taiwan}

\keywords{Quantum thermodynamics $|$  Quantum statistical mechanics $|$ Open quantum systems $|$ Exact master equation $|$ Nonequilibrium dynamics }

\begin{abstract}
By solving the exact master equation of open quantum systems, we formulate the quantum thermodynamics 
from weak to strong couplings. The open quantum systems exchange matters, energies and information with 
their reservoirs through quantum particles tunnelings that are described by the generalized Fano-Anderson Hamiltonians. 
We find that  the exact solution of the reduced density matrix of these systems approaches a Gibbs-type state in the 
steady-state limit for the systems in arbitrary initial states as well as for both the weak and strong system-reservoir coupling 
strengths. When the couplings become strong, thermodynamic quantities of the system must be renormalized. 
The renormalization effects are obtained nonperturbatively after exactly traced over all reservoir states through the 
coherent state path integrals. The renormalized system Hamiltonian is characterized by the renormalized system 
energy levels and interactions, corresponding to the quantum work done by the system.  The renormalized temperature 
is introduced to characterize the entropy production counting the heat transfer between the system and the reservoir.  
We further find that only with the renormalized system Hamiltonian and other renormalized thermodynamic 
quantities, can the exact steady state of the system be expressed as the standard Gibbs state. 
Consequently, the corresponding exact steady-state particle occupations in the renormalized 
system energy levels obey the Bose-Einstein and the Fermi-Dirac distributions for bosonic and 
 fermionic systems, respectively. In the very weak system-reservoir coupling limit, the renormalized 
 system Hamiltonian and the renormalized temperature are reduced to the original bare Hamiltonian of the system 
 and the initial temperature of the reservoir. Thus, the conventional statistical mechanics  
 and thermodynamics are thereby rigorously deduced from quantum dynamical evolution. In the last,
 this nonperturbative renormalization method is also extended to general interacting open quantum systems.
\end{abstract}

\date{This revised manuscript was compiled on Mar. 25, 2022}

\maketitle

\section{Introduction}
Understanding the physical process of thermalization within the framework of quantum mechanical principle has been a 
long-standing problem.  Thermodynamics and statistical mechanics are built with the hypothesis of equilibrium 
\cite{Landau1969,Kubo1991}, that is, over a sufficiently long time, a macroscopic system which is very weakly coupled 
with a thermal reservoir can always reach thermal equilibrium, and its equilibrium statistical distribution does not depend 
on the initial state of the system.  Over a century and a half, investigating the foundation of statistical mechanics and 
thermodynamics has been focused on two basic questions \cite{Huang1987}: (i) how does macroscopic irreversibility
emerge from microscopic reversibility? and (ii) how does the system relax to thermal equilibrium with its environment
from an arbitrary initial state? Rigorously solving these problems from the dynamical evolution of quantum systems, 
namely, finding the underlie of disorder and fluctuations from the deterministic dynamical evolution, has
 been a big challenge in physics \cite{Huang1987,Landau1969,Kubo1991,Feynman1963,Leggett1983a,Srednicki94,Zurek2003,Gemmer2004,Jarzynski2011,
Zhang2012,Kosloff2013,Nandkishore15,Xiong2015,MillenNJP2016,EspositoNJP17,Binder2018,Deffner2019,Xiong2020,Talkner2020}. 
Obviously, the foundation of thermodynamics and statistical mechanics and the 
answers to these questions rely on a deep understanding of the dynamics of systems interacting with their
environments, i.e., the nonequilibrium evolution of open quantum systems.

In 1980's,  Caldeira and Leggett investigated the problem of thermalization  from the study of the
quantum Brownian motion, a Brownian particle coupled to a thermal reservoir made by a continuous distribution of harmonic oscillators  
\cite{Leggett1983a}. They used the Feynman-Vernon influence functional approach \cite{Feynman1963} to explore the 
dynamics of quantum Brownian motion, and found  the equilibrium thermal state approximately \cite{Leggett1983a}.  
Later, Zurek studied extensively this nontrivial problem from the quantum-to-classical transition point of view. Zurek revealed 
the fact that thermalization is  realized through decoherence dynamics as a consequence of entanglement between  
the system and the reservoir \cite{Zurek2003}.  Thermalization in these investigations is demonstrated for quantum 
Brownian motion for initial Gaussian wave packets at high temperature limit \cite{Leggett1983a,Zurek2003}. 
However, the thermalization with arbitrary initial state of the system at arbitrary initial temperatures of one or 
multiple reservoirs for arbitrary system-reservoir coupling strengths have not been obtained. 

On the other hand, 
in the last two decades, experimental investigations on nano-scale quantum heat engines have 
attracted tremendous attentions on the realization of thermalization and the formulation of quantum thermodynamics  
\cite{Allahverdyan2000,Scully2003,Esposito2010a,Scully2011,Trotzky2012,Jezouin2013,KZhang2014,Bergenfeldt2014,Langen2015,
Pekola2015,David2015,Kaufman2016,Anders2016,Ronagel2016,Ochoa15}.
Besides searching new thermal phenomena arisen from quantum coherence and quantum entanglement,
an interesting question also appeared naturally is what happen when microscopic systems couple strongly with reservoirs. 
Since then, many effects have been devoted on the problems of how thermodynamic laws emerge
from quantum dynamics and how these laws may be changed when the system-reservoir couplings become strong
\cite{Campisi2009,Subas2012,Esposito2015,Seifert2016,Carrega2016,Ochoa2016,Jarzynski2017,Marcantoni2017,Bruch2018,
Perarnau2018,Hsiang2018,Anders2018,Strasberg2019,Newman2020,Ali2020a,Rivas2020}.
In particular, how to properly define the thermodynamic work and heat in the quantum mechanical framework
becomes an important issue when the system and reservoirs strongly coupled together 
\cite{Esposito2010b,Binder2015,Esposito2015,Alipour2016,Strasberg2017,Perarnau2018}. 
Due to various assumptions and approximations one inevitably taken in addressing the open quantum system dynamics,
no consensus has been reached in building quantum thermodynamics at strong coupling. 

In the last decade, we have derived the exact master equation of open quantum systems  
\cite{Tu2008,Jin2010,Lei2012,Yang2015,Yang2017,Zhang2018,Lai2018,Huang2020} by extending the Feynman-Vernon 
influence functional theory into the coherent state representation \cite{Zhang1990}. The open quantum systems
we have studied are a large class of nano-scale open quantum systems that exchange matters, energies and 
information with their reservoirs through the particle tunneling processes. We also solved the exact master 
equation of these systems with arbitrary initial states at arbitrary initial reservoir temperatures. Thus, a rather 
general picture of thermalization processes has been obtained \cite{Zhang2012,Xiong2015,Xiong2020}. 
In this paper, we shall explore the thermodynamic laws and statistical mechanics principles from the dynamical evolution of 
open quantum systems for both the weak and strong coupling strengths, based on the exact solution of the exact master 
equation we obtained. 

In fact, the difficulty for building the strong coupling quantum thermodynamics is twofold
\cite{Seifert2016,Carrega2016,Ochoa2016,Jarzynski2017,
Marcantoni2017,Bruch2018,Perarnau2018,Hsiang2018,Anders2018,Strasberg2019,Newman2020,Ali2020a,Rivas2020}:
(i) How to systematically determine the internal energy from the system Hamiltonian which
must be modified by the strong coupling with its reservoirs? (ii) How to correctly account the entropy
production when the system evolves from nonequilibrium state to the steady state?
We find that the nature of solving the above difficulty is the renormalization of both the system Hamiltonian
and the system density matrix during the nonequilibrium evolution through the system-reservoir couplings.
The system-reservoir couplings also result in the dissipation and fluctuation dynamics in open 
quantum systems, which are indeed renormalization effects of the system-reservoir interactions. 
The renormalization effects can be obtained nonperturbatively after exactly traced over all reservoir states. 
They are manifested in the dynamical evolution of the reduced density matrix 
with dissipation and fluctuation, and accompanied by the renormalized system Hamiltonian. 
We develop such a nonperturbative renormalization theory of quantum thermodynamics from weak to strong couplings
in this paper.

The rest of the paper is organized as follows. In Sec.~II, we begin with the simple open quantum system of a 
nanophotonic system coupled with a thermal reservoir. The renormalized system Hamiltonian is 
obtained in the derivation of the exact master equation for the reduced density matrix.
The exact solution of the reduced density matrix is also obtained analytically from the exact master equation.
Its steady state approaches to a Gibbs state so that quantum thermodynamics emerges naturally. 
However, we find that the exact solution of the particle occupation in the system at strong coupling 
does not agree to the Bose-Einstein distribution with the initial reservoir temperature. 
This indicates that the corresponding equilibrium temperature must also be renormalized when 
the reduced density matrix is influenced by the system-reservoir interaction through the dissipation and
fluctuation dynamics of the system.  By introducing the renormalized temperature as the derivative of 
the renormalized system energy with respect to the von Neumann entropy in terms of the reduced 
density matrix, we overcome the inconsistency. Thus, the self-consistent renormalized quantum statistics 
and renormalized quantum thermodynamics are formulated for both the weak and strong coupling strengths.

In Sec.~III, we extend such study to more general open quantum systems coupled to multi-reservoirs through particle 
exchange (tunneling) processes described by generalized Fano-Anderson Hamiltonians. 
These open systems are typical nano-scale systems that have been studied for various quantum 
transport in mesoscopic physics. Here both systems and reservoirs 
are made of many bosons or many fermions, not limiting to the prototypical open system of a harmonic oscillator 
coupling to a oscillator reservoir introduced originally by Feynman \cite{Feynman1963} and by Caldeira and Leggett 
\cite{Leggett1983a,Leggett1983b}. From the exact master equation and its exact solution in the steady state 
for such class of open quantum systems \cite{Tu2008,Jin2010,Lei2012}, we develop the renormalization theory 
of quantum thermodynamics for both the weak and strong coupling strengths in general. We further take an 
electronic junction system (a single electronic channel coupled two reservoirs with 
different initial temperatures and chemical potentials) as a specific nontrivial application. It is a nontrivial example 
because other approaches proposed for strong coupling quantum thermodynamics in the last few years
keep the reservoir temperature unchanged  \cite{Seifert2016,Carrega2016,Ochoa2016,Jarzynski2017,
Marcantoni2017,Bruch2018,Perarnau2018,Hsiang2018,Anders2018,Strasberg2019,Newman2020,Rivas2020}
so that these approaches become invalid for multiple reservoirs when the total system (the system plus all reservoirs) 
reaches a final equilibrium state.  We demonstrate the consistency of the Fermi-Dirac statistics with our renormalized 
quantum thermodynamic in this nontrivial application.

In Sec.~IV, we discuss further the generalization of this nonperturbative renormalization theory for quantum 
thermodynamics to more complicated interacting open quantum systems.
We take the non-relativistic quantum electrodynamics (QED) derived from the fundamental quantum field theory 
as an example, and considered electrons as the open system and all photonic modes (electromagnetic field) 
as the reservoir. The system-reservoir interaction is the fundamental electron-photon interaction. 
We perform the nonperturbative renormalization by integrating out exactly the infinite number of 
electromagnetic field degrees of freedom. We obtain the reduced density matrix in terms of only the system 
degrees of freedom in the same way as we derived the exact master equation for the generalized 
Fano-Anderson Hamiltonian in Sec.~III.  The resulting renormalization theory are given  
by the reduced density matrix for electrons and the nonperturbative renormalized electron Hamiltonian,
which can be systemically computed in terms of two-electron propagating Green functions in principle.
Thus, we show that although our renormalized quantum thermodynamics theory is formulated from the exact 
solvable open quantum systems, it can apply to arbitrary open quantum systems even though the final exact 
analytical solution is hardly found. In fact, similar situation also exists for the equilibrium statistical mechanics, 
namely, one cannot solve exactly 
all equilibrium physical systems, in particular the strongly correlated systems such as the Hubbard model and the 
general quantum Heisenberg spin model \cite{Nagaosa2010} even though the reservoir effect can be ignored there.  
Therefore, approximations and numerical methods remained to be developed further for the study of renormalized 
nonequilibrium dynamics within the framework we developed  in this paper. 
A conclusion is given in Sec.~V. In Appendices, we provide the necessary analytical derivations of 
the solutions used in the paper.  

\section{A simple example for strong coupling quantum thermodynamics}
For simplicity, we begin with a single-mode bosonic open system (such as a microwave cavity in quantum optics
or a vibrational phononic mode in solid-state and biological systems) coupled  to a thermal reservoir
through the energy exchange interaction. 
The total Hamiltonian of the system, the reservoir and the coupling between them is considered to be described by the
Fano Hamiltonian \cite{Fano1961}:
\begin{align}
H_{\rm tot}\!= & H_\s\!+\!H_\e\!+\!H_{\s\e} \notag \\
=& \hbar \w_s a^\+a\!+\!{\sum}_k\hbar\w_kb^\+_kb_k\!+\!{\sum}_k\hbar(V_ka^\+b_k\!+\!V_k^*b^\+_ka), \label{fH}
\end{align}
where $a^\+$ and $b^\+_k$ ($a$ and $b_k$) are the creation (annihilation) operators of the bosonic modes in the
system and in the reservoir with energy quanta $\hbar\w_s$ and $\hbar\w_k$, respectively. They obey the standard 
bosonic commutation relations: $[a, a^\dag]=1$ and $[b_k,b^\dag_{k'}]=\delta_{kk'}$, etc. The parameter
 $V_k$ is the coupling amplitude between the system and the reservoir and can be experimentally tuned 
 to strong coupling \cite{Putz14,Chiang21}. 
 In fact, all parameters in the Hamiltonian, including the couplings between the system and the reservoir can be 
 time-dependently controlled with the modern nano and quantum technologies. The universality of Fano resonance also 
 makes this simple system useful in nuclear, atomic, molecular and optical physics, as well as condensed matter systems \cite{Miroshnichenko2010}.

\subsection{The exact master equation of the system and its exact nonequilibrium solution}
To study the thermalization of open quantum systems, the reservoir can be initially set in a thermal state 
\begin{align}
\rho_\e(t_0)\!=\!e^{-\beta_0 H_\e}/Z_\e,
\end{align} 
where $\beta_0\!=\!1/k_B T_0$ and
$T_0$ is the initial temperature of the reservoir, $Z_\e\!=\!\Tr_\e[e^{-\beta_0 H_\e}]$ is its
partition function. The system can be initially in arbitrary state $\rho_\s(t_0)$ so that the initial total density matrix of the 
system plus the reservoir is a direct product state \cite{Feynman1963,Leggett1983a},
\begin{align}
\rho_{\rm tot}(t_0)= \rho_\s(t_0)\otimes \f{\!e^{-\beta_0 H_\e}}{Z_\e}.  \label{itdm}
\end{align} 
After the initial time $t_0$, both the system and the reservoir evolve into an entangled nonequilibrium state 
$\rho_{\rm tot}(t)$ which obeys the Liouville-von Neumann equation in quantum mechanics \cite{Neumann55},
\begin{align}
\frac{d}{dt}\rho_{\rm tot}(t)=\frac{1}{i\hbar}[H_{\rm tot}, \rho_{\rm tot}(t)].   \label{voneq}
\end{align} 
Because the system and the reservoir together form a closed system, the Liouville-von Neumann equation is the same as 
the Schr\"{o}dinger equation of quantum mechanics for the evolution of pure quantum states. But the Liouville-von Neumann 
equation is more general because it is also valid for mixed states. 

Quantum states of the system are completely determined by the reduced density matrix $\rho_\s(t)$. It is defined 
by the partial trace over all the reservoir states: 
\begin{align}
\rho_\s(t)\!=\!\Tr_\e[\rho_{\rm tot}(t)]. 
\end{align}
The equation of motion for $\rho_\s(t)$, which 
is called the master equation, determines the quantum evolution of the system at later time $t$ ($>t_0$). 
In the literature, one usually derives the master equation using various approximations, such as the 
memory-less dynamical maps, the Born-Markovian approximation, and secular approximation, etc.~\cite{Lindblad1976,GKS1976,Kolodynski2018,Breuer2008,Paavola2009,Rajesh2015}. 
But these methods are invalid for strong coupling open quantum systems with strong non-Markovian dynamics.  
In the past decade, we have developed a very different approach 
to rigorously derive the exact master equation for a large class of open quantum systems \cite{Tu2008,Jin2010,Lei2012,Yang2015,Yang2017,Lai2018,Zhang2018,Huang2020}.
Explicitly, we have derived the exact master equation for Eq.~(\ref{fH}) by exactly tracing over all the reservoir 
states from the solution of the Liouville-von Neumann equation \cite{Wu2010,Xiong2010,Lei2011,Lei2012}. 
The trace over all the reservoir states is a nonperturbative renormalization to the reduced density matrix of the system 
and to the system Hamiltonian simultaneously. We complete this partial trace by integrating out exactly all the reservoir 
degrees of freedom through the coherent state path integrals \cite{Lei2012,Zhang1990}. 
The resulting exact master equation for the reduced density matrix accompanied with the renormalized system Hamiltonian is given by
\begin{subequations}
\label{emefh}
\begin{align}
  \f{d}{dt}\rho_\s & (t)\!= \frac{1}{i\hbar}\big[ H^r_\s(t),\rho_\s(t)\big]  \notag\\
  & \!+\!\gamma(t,t_0)\big\{2a\rho_\s(t)a^\+ \!-\! a^\+a\rho_\s(t) \! -\! \rho_\s(t)a^\+a\big\}  \notag\\
  &  \!+\! \widetilde{\gamma}(t,t_0)\big\{a^\+ \rho_\s(t)a \!+\! a\rho_\s(t)a^\+\!-\! a^\+a\rho_\s(t) \!-\! \rho_\s(t)aa^\+\big\}.  \label{eme1}
\end{align}
In this exact master equation, the first term describes a unitary evolution of the reduced density matrix with the renormalized 
Hamiltonian 
\begin{align}
H^r_\s(t)=\hbar \w^r_s(t,t_0)a^\+a .  \label{rsH}
\end{align}
\end{subequations}
This renormalized Hamiltonian contains all the energy corrections to the system arisen from the system-reservoir interaction 
through the nonequilibrium evolution. The second and 
the third terms describe the non-unitary evolution of the reduced density matrix, which contain all non-Markovian dissipation and 
fluctuation dynamics induced by the back-reactions between the system and the reservoir through the system-reservoir interaction
(see Eqs.~(\ref{fmrc}) and (\ref{uvte}) given later).
Physically, the second and the third terms in the above master equation also characterize the emergence of disorder and fluctuations 
induced by the system-reservoir interaction. This is because if the system is initially in a pure quantum state, it contains zero disorder 
at beginning (its initial entropy is zero). The index $r$ denotes renormalized physical quantities hereafter.  

The energy renormalization, the dissipation and fluctuation dynamics described in the exact master equation Eq.~(\ref{emefh})
are characterized by the non-Markovian renormalized frequency $\w^r_s(t,t_0)$, the non-Markovian dissipation coefficient 
$\gamma(t,t_0)$ and the non-Markovian fluctuation coefficient $\widetilde{\gamma}(t,t_0)$, respectively. 
All these non-Markovian coefficients are nonperturbatively and exactly determined 
by the following relations \cite{Wu2010,Xiong2010,Lei2011,Lei2012}
\begin{subequations}
\label{fmrc}
\begin{align}
 & \w^r_s(t,t_0) = - {\rm Im}[\dot{u}(t,t_0)/u(t,t_0)],  \label{re} \\
 & \gamma(t,t_0)  = - {\rm Re}[\dot{u}(t,t_0)/u(t,t_0)], \label{ds} \\
 & \tilde{\gamma}(t,t_0)= \dot{v}(t,t)-2v(t,t){\rm Re}[\dot{u}(t,t_0)/u(t,t_0)] .
\end{align}
\end{subequations}
Here $u(t,t_0)$ and $v(\tau,t)$ are the two non-equilibrium Green functions obeying the integro-differential Dyson equations, 
\begin{subequations}
  \label{uvte}
\begin{align}
  & \f{d}{dt}u(t,t_0) \!+\! i\w_s u(t,t_0) \!+ \!\! \int^t_{t_0} \!\! d\tau g(t,\tau) u(\tau,t_0)=0 , \label{ut} \\
  & v(\tau,t)= \! \int^\tau_{t_0} \!\! d\tau_1\! \int^t_{t_0} \!\! d\tau_2 u(\tau,\tau_1)\widetilde{g}(\tau_1 , \tau_2)u^*(t,\tau_2).
\end{align}
\end{subequations}
The non-Markovianity is manifested by the above time-convolution equation of motion for these non-equilibrium Green functions.
The integral kernels in the above convolution equations are given by 
\begin{subequations}
\begin{align}
&g(t,\tau)=\!\! \int^\8_0 \!\!\! d\w J(\w)e^{-i\w(t-\tau)},  \label{gt} \\
&\widetilde{g}(\tau_1,\tau_2)=\!\! \int^\8_0 \!\!\! d\w J(\w) \overline{n}(\w,T_0)e^{-i\w(\tau_1-\tau_2)},
\end{align}
\end{subequations}
which characterize the time correlations between the system and the reservoir through the system-reservoir interaction. 
The frequency-dependent function 
\begin{align}
J(\w)\equiv \sum_k|V_k|^2\delta(\w-\w_k)   \label{sd}
\end{align} 
is called as the spectral density, 
which fully encapsulates the fundamental dissipation (relaxation) and fluctuation (noise or dephasing) effects induced by the
system-reservoir interaction. Finally, the initial temperature dependent function 
\begin{align}
\overline{n}(\w_k,T_0)=\!\Tr_\e[b^\dag_kb_k \rho_\e(t_0)]= 1/[e^{\hbar \w_k/k_BT_0}-1]
\end{align} 
is the initial particle distribution in the reservoir. 

An arbitrary initial state of the system can be expressed as 
\begin{align}
\rho_\s(t_0)\!=\!\sum^\8_{m,m'=0}\rho_{mm'}|m\ket \bra m'|,  \label{inis}
\end{align}
where $|m\rangle= \frac{1}{\sqrt{m!}}(a^\dag)^m|0\rangle$ is the bosonic Fock state. If  
$\rho_{mm'}=c_m c^*_{m'}$, then $\rho_\s(t_0)$ is a pure state, otherwise it is a mixed state.
The exact solution of the exact master equation Eq.~(\ref{eme1}) has be found \cite{Xiong2015,Xiong2020},
  \begin{align}
  \rho^{\rm exact}_\s(t)=&\! \!\!\! \sum^\8_{m,m'=0} \!\!\! \rho_{mm'} \!\!\!\!\!\!\! \sum_{k=0}^{\rm min\{m,m'\}} 
  \!\!\!\!\!\!\! d_k(t) A^\+_{mk}(t)\widetilde{\rho}[v(t,t)] A_{m'k}(t),   \label{esrdm}
  \end{align}
  where 
  \begin{subequations}
  \begin{align}
&  \widetilde{\rho}[v(t,t)]=\sum_{n=0}^\8\f{[v(t,t)]^n}{[1+v(t,t)]^{n+1}}|n\ket\bra n|, \\
& A^\+_{mk}(t)=\f{\sqrt{m!}}{(m-k)!\sqrt{k!}}\Big[\frac{u(t,t_0)}{1+v(t,t)}a^\+\Big]^{m-k}, \\
& d_k(t)=\!\Big[1\!-\!\f{|u(t,t_0)|^2}{1+v(t,t)}\Big]^k.
 \end{align}
 \end{subequations}
As a self-consistent check of the above solution, we calculate the average particle number in the system from the above solution,
$\overline{n}(t)\equiv\Tr_\s[a^\+a\rho_\s(t)]$, also using the Heisenberg equation of motion directly,
$\overline{n}(t)\equiv\Tr_{\s+\e}[a^\+(t)a(t)\rho_{\rm tot}(t_0)]$. Both calculations give the same result 
\cite{Jin2010,Lei2012,Yang2015,Zhang2018}:
\begin{align}
\overline{n}(t)& =\Tr_\s[a^\+a\rho_\s(t)]=\Tr_{\s+\e}[a^\+(t)a(t)\rho_{\rm tot}(t_0)] \notag \\
& = u^*(t,t_0) \overline{n}(t_0) u(t.t_0) + v(t,t),  \label{pnexp}
\end{align}
where $u(t,t_0)$ and $v(t,t)$ are determined by Eq.~(\ref{uvte}).

Based on the above exact formalism, for a given spectral density $J(\w)$,  if no localized bound state exists \cite{Zhang2012,Expl},
the general solution of Eq.~(\ref{ut}) is 
\begin{align}
u(t,t_0) = \int_0^{\8}\!\!\!\! d\omega D(\w)e^{i\omega(t-t_0)} \xrightarrow{t\ra\8} 0  \label{ut0}
\end{align} 
where $ D(\w) \! = \! \f{J(\w)}{[\w-\w_s-\Delta(\w)]^2 + \pi^2J^2(\w)}$ shows the system spectrum broadening due to the coupling to the reservoir,
 and $\Delta(\w) = {\cal P}\big[\! \int d\w' \f{J(\w')}{\w-\w'}\big]$ is the principal-value integral of the self-energy correction to the system, 
 $\Sigma(\omega)\!=\! \int d\omega' \frac{J(\omega')}{\omega-\omega'}\!=\!\Delta(\omega)-i\pi J(\omega)$. In fact, $\Delta(\w)$ 
 gives the system frequency (or energy) shift.
 As a result, in the steady state limit, we have $A^\+_{mk}(t) \xrightarrow{t\ra\8} \delta_{mk}$ and $d_k(t)\xrightarrow{t\ra\8}1$. Then 
the exact solution of the particle distribution and the reduced density matrix  are reduced to
\begin{subequations}
\label{smss}
\begin{align}
  \overline{n}_{\rm exact}\!(t\!\ra\!\8) &= \lim_{t\ra\8}v(t,t) = \!\! \int^\infty_0 \!\! \!\! d\w D(\w)\overline{n}(\w,T_0),  \label{sspd} \\
  \rho^{\rm exact}_\s\!(t\!\ra\!\8) &= \lim_{t\ra\8}\sum_{n=0}^\8 \f{[v(t,t)]^n} {[1+ v(t,t)]^{n+1}} | n \ket \bra n |  \\
&=\lim_{t\ra\8} \f{\exp\big\{\ln\!\big[\!\f{v(t,t)}{1+v(t,t)}\big]a^\+a\big\}}{1+v(t,t)} .  \label{rss}
 \end{align}
 \end{subequations}
 Equation (\ref{smss}) is the exact steady-state solution of the system coupled to a thermal reservoir for all 
 coupling strengths for the open system Eq.~(\ref{fH}).  All the influences of the reservoir on the system through the 
 system-reservoir interaction have been taken into account in this solution.
 Remarkably, the above results show that the exact solution of the steady state is independent of the initial state 
 of the system and is determined by the particle distribution, as a consequence of thermalization \cite{Xiong2015,Xiong2020}.
 
Note that the above exact master equation formalism remains the same for initial states involved initial correlations 
between the system and the reservoir, with the only modification of the correlation function $v(\tau,t)$, as we have 
shown in Refs.~\cite{Tan2011,Huang2020,Yang2015}. 
This exact master equation formalism has also been extended to open quantum systems including external deriving fields 
\cite{Lei2012,Chiang21}.

\subsection{Renormalization of quantum thermodynamics}
\begin{figure}[ht]
\centering
\includegraphics[width=5.50cm]{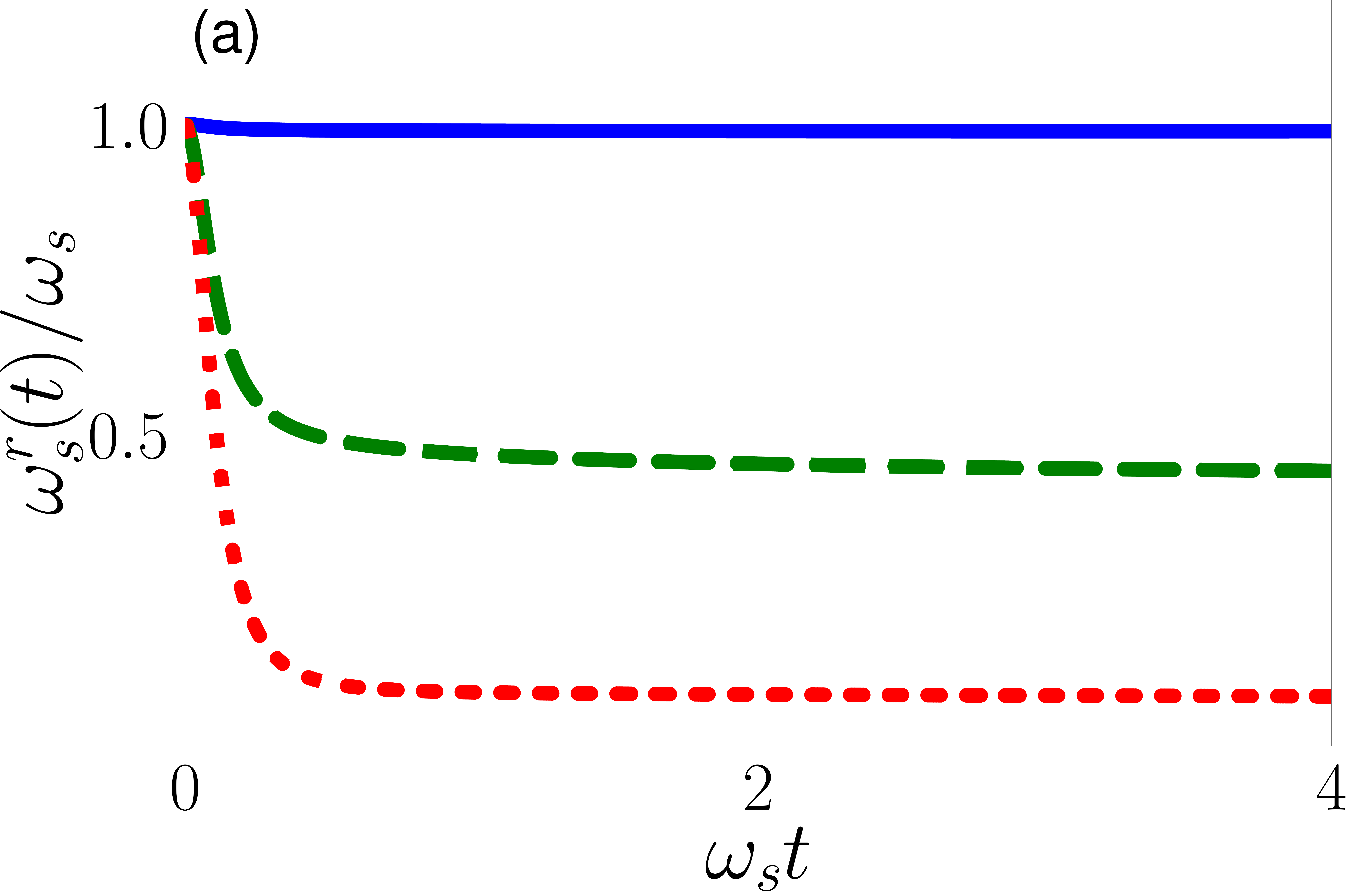}
\includegraphics[width=5.5cm]{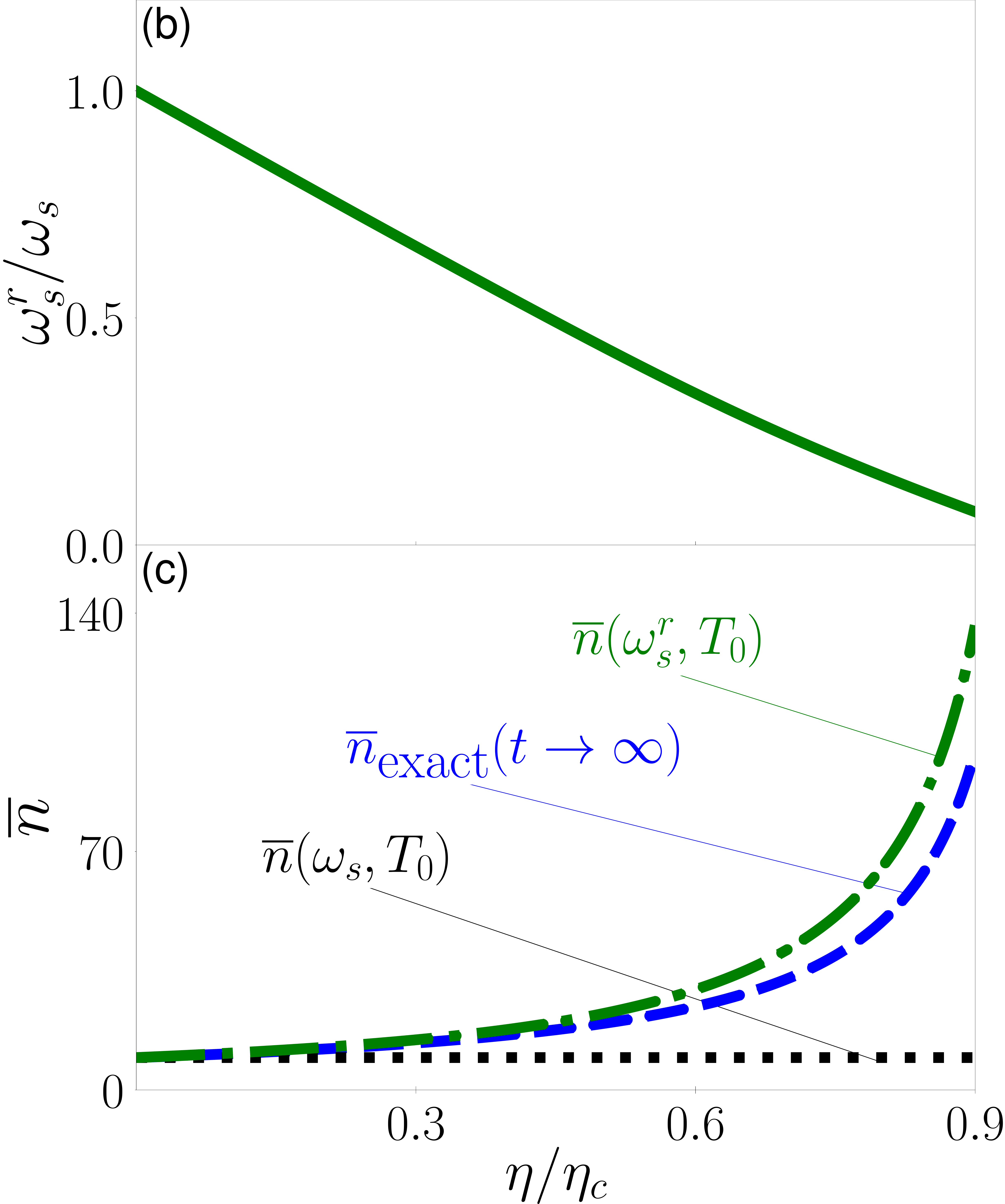}
\caption{(a) The renormalized system energy (cavity frequency shift) $\hbar\omega^r_s(t)$ for three different system-reservoir 
coupling strengths, $\eta=0.01\eta_c, 0.5 \eta_c, 0.9 \eta_c$. It is calculated from Eqs.~(\ref{re}) and (\ref{ut}) for the Ohmic spectral density 
$J(\w)\!=\!\eta\w\exp(-\w/\w_c)$, where the cutoff frequency  $\omega_c=10\omega_s$ is taken, and
$\eta_c\!=\!\omega_s/\omega_c$ is a critical coupling for the Ohmic spectral density 
\cite{Zhang2012,Xiong2015}.
(b) The steady-state renormalized frequency shift $\omega^r_s =\omega^r_s (t\!\ra\!\8)$ as a function 
of the system-reservoir coupling strength $\eta/\eta_c$.
(c) The steady-state particle distribution $\overline{n}_{\rm exact}(t\!\ra\!\8)$ of Eq.~(\ref{sspd}) (the blue-dashed line),
the Bose-Einstein distribution without the energy (frequency) renormalization   $\overline{n}(\w_s,T_0)$ (the black-dot line) 
and with the energy renormalization  $\overline{n}(\w^r_s,T_0)$ (the green-dashed-dot line), respectively.
The system is initially set in a pure Fock state $|n_0\rangle$ with $n_0=5$, and the reservoir initial temperature
$T_0=10 \hbar \omega_s$.
\label{pdsc}}
\end{figure}

Now we can study quantum thermodynamics for all the coupling strengths from the above exact solution.  
First, the master equation Eq.~(\ref{emefh}) shows that the Hamiltonian of the system must be renormalized
from $H_\s$ to $H^r_\s$ given by the energy (or frequency) shift from $\hbar\w_s$ to $\hbar\w^r_s(t)$ during 
the nonequilibrium dynamical evolution. This is a nonperturbative renormalization effect of the system-reservoir coupling on the system. 
The renormalized frequency $\w^r_s(t)$ and its steady-state value $\w^r_s=\w^r_s(t\!\ra\!\8)$ can be exactly calculated from
Eqs.~(\ref{re}) and (\ref{ut}). Here we take the Ohmic spectral density $J(\w)\!=\!\eta\w\exp(-\w/\w_c)$
\cite{Leggett1987} in the practical calculation. The result is presented in Fig.~\ref{pdsc}(a) and (b). It shows that different 
system-reservoir coupling strengths $\eta$ will cause different renormalized system energies, resulting in
different cavity frequency shifts, see Fig.~\ref{pdsc}(a).  In Fig.~\ref{pdsc}(b), we plot the steady-state values of the renormalized cavity 
frequency as a function of the system-reservoir coupling strength $\eta/\eta_c$, where $\eta_c=\w_s/\w_c$ is a 
critical coupling strength for the Ohmic spectral density \cite{Zhang2012,Xiong2015}. 
When $\eta >\eta_c$, the system-reservoir coupling would generate a localized mode (localized bound state)
such that the cavity system cannot approach to the equilibrium with the reservoir, as we will discuss later \cite{Xiong2015,Xiong2020}.

In Fig.~\ref{pdsc}(c), we plot the exact solution $\overline{n}_{\rm exact}(t\!\ra\!\8)$ of Eq.~(\ref{sspd}) 
as a function of the coupling strength $\eta/\eta_c$ (the blue-dashed line). We compare the result with the Bose-Einstein 
distribution without the energy (frequency) renormalization,  $\overline{n}(\w_s,T_0)=1/[e^{\hbar \w_s/k_BT_0}-1]$ 
(see the black-dot line),  also compare to Bose-Einstein distribution with the energy renormalization,  
$\overline{n}(\w^r_s,T_0)=1/[e^{\hbar \w^r_s/k_BT_0}-1]$ (see the green-dashed-dot line).  
As one can see, the exact solution $\overline{n}_{\rm exact}(t\!\ra\!\8)$ derivates significantly from the Bose-Einstein distribution 
without the energy renormalization, i.e.,~$\overline{n}(\w_s,T_0)$, as $\eta$ increases.  This derivation shows how the system-reservoir 
coupling strength changes the intrinsic thermal property of the system. On the other hand, the Bose-Einstein distribution 
with the renormalized energy, given by $\overline{n}(\w^r_s,T_0)$, changes with the changes of $\eta$, similar to the exact solution 
$\overline{n}_{\rm exact}(t\!\ra\!\8)$. But there is still a quantitative disagreement between the exact solution $\overline{n}_{\rm exact}(t\!\ra\!\8)$
and the Bose-Einstein distribution $\overline{n}(\w^r_s,T_0)$ with the renormalized cavity photon energy $\hbar\w^r_s$.

To understand further the origin of the above difference, let us recall that the exact solution 
$\rho^{\rm exact}_\s\!(t\!\ra\!\8)$ of Eq.~(\ref{rss}) is indeed a Gibbs-type state.
This indicates that the exact particle distribution $\overline{n}_{\rm exact}(t\!\ra\!\8)$ should obey a 
Bose-Einstein distribution for all coupling strengths. To find such distribution that agrees with the exact solution
Eq.~(\ref{sspd}), one possibility is to renormalize the temperature, because no other thermal quantity
can be modified in the Gibbs state for this photonic system. 
In the literature, it is commonly believed that the reservoir is large enough so that its temperature 
should keep invariant \cite{Seifert2016,Jarzynski2017}. 
However,  the initial decoupled states Eq.~(\ref{itdm}) of the system plus the reservoir is not an equilibrium state 
of the total system. After the initial time $t_0$,
both the system and reservoir evolve into a correlated (entangled) nonequilibrium state $\rho_{\rm tot}(t)$. 
When the system and the reservoir reach the equilibrium state, there must be a fundamental way to show whether the
new equilibrium state is still characterized by the initial reservoir temperature. 

As a self-consistent check, let us denote 
the final steady-state equilibrium temperature as $T_f$. Then, according to the equilibrium statistical mechanics,  
the steady state of the total system (the system plus the reservoir) should be 
\begin{align}
\rho_{\rm tot}(t\rightarrow \infty) = \frac{1}{Z_{\rm tot}} e^{-\beta_f H_{\rm tot}} , \label{tsdm}
\end{align}
where $\beta_f=1/k_BT_f$, and $H_{\rm tot}$ is the total Hamiltonian of the system plus the reservoir, including the 
coupling interaction between them, i.e.,~Eq.~(\ref{fH}).
Taking a trace over the reservoir states from the above steady state of the total density matrix, we have rigorously proven 
\cite{Huang2020} that (also see the detailed derivation given in Appendix A)
\begin{align}
\rho_\s(t\rightarrow \infty) \! & =\! \Tr_\e\Big[\frac{e^{-\beta^r H_{\rm tot}}}{Z_{\rm tot}}\Big] \notag \\
&=\! \f{\exp\big\{\ln\!\big[\!\f{\overline{n}(t\!\ra\!\8)}{1+\overline{n}(t\!\ra\!\8)}\big]a^\+a\big\}}{1+\overline{n}(t\!\ra\!\8)}.
 \label{ess}
 \end{align}
This result is the same as the solution of Eq.~(\ref{smss}). The latter is the steady state of the exact time-dependent solution 
Eq.~(\ref{esrdm}) solved  from the exact master equation Eq.~(\ref{eme1}) for arbitrary coupling. This shows that 
the equilibrium state Eq.~(\ref{tsdm}) which is originally proposed in statistical mechanics 
is indeed valid for both the weak and strong coupling between the system 
and the reservoir. Furthermore, the exact particle distribution can be obtained from the dynamical evolution of exact master equation 
or from the Heisenberg equation of motion directly, as shown by Eq.~(\ref{pnexp}). Thus, we have
$\overline{n}(t\!\ra\!\8)= \Tr_{\s+\e}[a^\dag a \rho_{\rm tot}(t\!\ra\!\8)]= \Tr_\s[a^\dag a \rho_\s(t\!\ra\!\8)]
=\overline{n}_{\rm exact}(t\!\ra\!\8)$. This gives a further self-consist justification to the above conclusion.

The result presented in Fig.~\ref{pdsc}(c) shows that $\overline{n}(\w^r_s,T_0) \neq  \overline{n}_{\rm exact}(t\!\ra\!\8)$ 
except for the weak coupling. This indicates that in general, $T_f \neq T_0$, namely the final equilibrium temperature of the total
system cannot be the same as the initial equilibrium temperature of the reservoir when the total system reaches the new 
equilibrium state, except for the very weak coupling strength.
Now the question is how to determine this steady-state equilibrium temperature $T_f$ when the system and the reservoir finally reach  
the equilibrium state. According to the axiomatic description of thermodynamics \cite{Callen1985},  the equilibrium temperature
of a system is defined as the change of its internal energy with respect to the change of its thermal entropy.  
This temperature definition in thermodynamics does not assume a weak coupling between the system and 
the reservoir because no statistical mechanics is used in this definition. It is the fundamental definition of the temperature for arbitrary 
two coupled thermodynamic systems when they reach the equilibrium each other, from which the zeroth law of thermodynamics is derived 
\cite{Callen1985}. 

Now, the average energy of the system at arbitrary time, i.e., the nonequilibrium internal energy of the system, is given by the renormalized 
Hamiltonian Eq.~(\ref{rsH}) with the exact solution of the reduced density matrix $\rho_\s(t)$ of Eq.~(\ref{esrdm}):
\begin{align}
U_\s(t) \equiv \Tr_\s[H^r_\s\!(t)\rho_\s(t)].   \label{intE}
\end{align}
Also, we define the von Neumann entropy with the exact reduced density matrix of Eq.~(\ref{esrdm}) as 
the nonequilibrium thermodynamic entropy of the system \cite{Ali2020a,Neumann55,Callen1985}:
\begin{align}
S_\s(t)=\!- k_B\Tr_\s[\rho_\s(t)\ln\rho_\s(t)],  \label{Entr}
\end{align}
where $k_B$ is the Boltzmann constant. Note that this entropy is defined for the exact reduced density matrix obtained
after traced over exactly all the reservoir states. It also encapsulates all the renormalization 
effects of the system-reservoir interaction to the system state distributions. Thus, we introduce a renormalized 
nonequilibrium thermodynamic temperature \cite{Ali2020} which is defined as
\begin{align}
 T^r\!(t) \equiv  \f{\pd U_\s(t)}{\pd S_\s(t)}\bigg|_{\omega^r_s} \!\! = \Tr_\s\bigg[H^r_\s\!(t) \f{d\rho_\s(t)}{dS_\s(t)}\bigg] . \label{rdT}
\end{align}
This is a direct generalization of the concept of the equilibrium temperature to nonequilibrium states in open quantum systems. 
When the system and its reservoir reach the equilibrium steady state, no matter the system-reservoir coupling is strong or weak, 
we can fundamentally obtain the final equilibrium temperature 
$T_f \equiv T^r\!=\!T^r(t\!\ra\!\8)$ from the dynamical evolution of the open quantum system. 

\begin{figure}[ht]
\centering
\includegraphics[width=8.5cm]{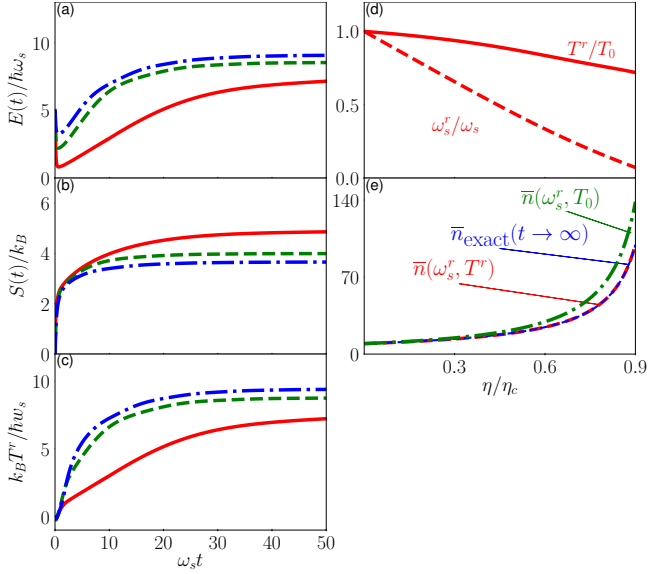}
\caption{ 
(a)-(c) The nonequilibrium dynamical evolution of the internal energy, the entropy and the corresponding renormalized 
temperature, given respectively by Eqs.~(\ref{intE})-(\ref{rdT}) 
for different coupling strengths $\eta / \eta_c = 0.3, 0.5, 0.8$ 
(correspond to the blue-dot-dashed, green-dashed, red-dot lines, respectively). 
(d) The steady-state values of the renormalized frequency $\w^r_s$ and renormalized temperature $T^r$ as 
a function of the coupling strength $\eta$.
(e) The steady-state particle distribution as a function of coupling strength $\eta$. It shows that the
exact solution $\overline{n}_{\rm exact}(t\!\ra\!\8)$ of Eq.~(\ref{sspd}) (the blue-dashed line) and 
the Bose-Einstein distribution $\overline{n}(\w^r_s,T^r)$ with the renormalized frequency 
and the renormalized temperature (the red-dot line) agree perfectly to each other. 
The green-dashed-dot line is  $\overline{n}(\w^r_s,T_0)$ without the renormalized temperature, 
which cannot describe the exact solution solved from the exact master equation.
Other parameters are taken as the same as that in Fig.~\ref{pdsc}. 
\label{pdsc2}}
\end{figure}

From the exact solution of the reduced density matrix $\rho^{\rm exact}_\s(t)$ of Eq.~(\ref{esrdm}), we calculate 
the time-dependence of the internal energy $U_\s(t)$, the entropy $S_\s(t)$ and the dynamical renormalized 
temperature $T^r(t)$ for different
coupling strengths. The corresponding results are presented in Fig.~\ref{pdsc2}(a)-(c), respectively. It shows 
explicitly how the nonequilibrium internal energy, entropy and renormalized temperature evolve 
differently for different system-reservoir coupling strengths. Their steady-state values also approach different points
for different coupling strengths. The different steady-state internal energies and entropies associated with different 
system-reservoir coupling strengths result in different 
steady-state temperatures. This indicates that the reservoir cannot remain unchanged from the initial reservoir temperature.
This new feature has not been discovered or noticed  in all previous investigations of strong-coupling quantum thermodynamics
\cite{Seifert2016,Carrega2016,Ochoa2016,Jarzynski2017,Marcantoni2017,Bruch2018,
Perarnau2018,Hsiang2018,Anders2018,Strasberg2019,Newman2020,Rivas2020}. 

In Fig.~\ref{pdsc2}(d), we plot the steady-state
 renormalized temperature, $T^r\!=\!T^r(t\!\ra\!\8)$, as a function of the coupling astrength $\eta/\eta_c$.
Using this renormalized temperature, we further plot the Bose-Einstein distribution with both the renormalized energy and 
the renormalized temperature: $\overline{n}(\w^r_s,T^r)=1/[e^{\hbar \w^r_s/k_BT^r}-1]$,  see the red-dot line
in Fig.~\ref{pdsc2}(e). \textit{Remarkably, it perfectly reproduces the exact solution of Eq.~(\ref{sspd}),} i.e., 
\begin{align}
\overline{n}_{\rm exact}(t\!\ra\!\8)=\overline{n}(\w^r_s,T^r)=\frac{1}{e^{\hbar \w^r_s/k_BT^r}-1}.
\end{align} 
In other words, in the steady state, the exact solution of the steady-state particle occupation 
solved from the exact dynamics of the open quantum system obeys the standard Bose-Einstein distribution
only for the renormalized Hamiltonian Eq.~(\ref{rsH}) with the renormalized temperature Eq.~(\ref{rdT}). 
This provides a very strong proof that strong coupling quantum thermodynamics must be renormalized
for both the system Hamiltonian and the temperature.

Furthermore, in terms of the renormalized Hamiltonian Eq.~(\ref{rsH})  and the renormalized temperature Eq.~(\ref{rdT}),  
the steady state Eq.~(\ref{rss}) can be expressed as the standard Gibbs state,
\begin{align}
\rho^{\rm exact}_\s\!(t\!\ra\!\8)\! & = \!\! \sum_{n=0}^\8 \!\f{[\overline{n}(\w^r_s,T^r)]^n} {[1+ \overline{n}(\w^r_s,T^r)]^{n+1}} 
| n \ket \bra n | \notag \\ & =\! \frac{1}{Z^r_\s} e^{-\beta^r\!H^r_s} ,  \label{rsspd}
\end{align}
where 
$Z^r_\s\!=\!\Tr_\s[e^{-\beta^r H^r_\s}]$ is the renormalized partition function, and $\beta^r\!=\!1/k_BT^r$ 
is the inverse renormalized temperature in the steady state.  This is a direct proof 
of how the statistical mechanics, as a consequence of disorder or randomness in the nature, 
emerges from the exact dynamical evolution of quantum mechanics. 

Moreover, one can check that in the very weak
coupling regime $\eta\!\ll\!\eta_c$, $\Delta(\w)\!\ra\!0$ and $D(\w)\!\ra\!\delta(\w-\w_s)$
 so that the steady state solution of Eq.~[\ref{sspd}] is directly reduced to $\overline{n}(\w_s,T_0)$
 \cite{Xiong2015,Xiong2020},
and
\begin{align}
 \rho^{\rm exact}_\s\!(t\!\ra\!\8)\! \xrightarrow{\eta\ll \eta_c} & \sum_{n=0}^\8\! \f{[\overline{n}(\w_s,T_0)]^n} 
 {[1\!+\!\overline{n}(\w_s,T_0)]^{n+1}} | n \ket \bra n | \notag \\
  & =\! \frac{1}{Z_\s} e^{-\beta_0 H_\s} .  \label{rssw}
 \end{align}
 This reproduces the expected solution of the statistical mechanics in the weak coupling regime.
 Figures \ref{pdsc} and \ref{pdsc2} also show that $\hbar\w^r_s\!\ra\!\hbar\w_s$ and $T^r\!\ra\!T_0$
at very weak coupling. Thus, the equilibrium hypothesis of thermodynamics and statistical mechanics
is deduced rigorously from the dynamics of quantum systems. This solves the long-standing problem
of how thermodynamics and statistical mechanics emergence from quantum dynamical evolution \cite{Huang1987}.

On the other hand, $\eta_c= \omega_s/\omega_c$ is a critical coupling strength for Ohmic spectral density that
when $\eta > \eta_c$, the system exists a dissipationless localized bound state (localized mode) at frequency 
$\omega_b=\omega_s + \Delta(\omega_b)$ with $J(\omega_b)=0$ \cite{Zhang2012}. 
Once such a localized mode exists, the spectral function $D(\omega)$ of the system in Eq.~(\ref{ut0}) is modified as
\begin{align}
 D(\w) \! = Z(\omega_b) \delta(\omega\!-\!\omega_b) + \! \f{J(\w)}{[\w\!-\!\w_s\!-\!\Delta(\w)]^2 \!+\! \pi^2J^2(\w)},
 \end{align}
 where $Z (\omega_b) = [1-\partial \Sigma(\omega)/\partial \omega]^{-1}\big|_{\w=\w_b}$ is the localized bound state wavefunction. 
Then, the asymptotic value of the Green function $u(t\!\ra\!\8,t_0)$ never vanishes. As a result, 
the steady state of the reduced density matrix Eq.~(\ref{esrdm}) cannot be reduced to Eq.~(\ref{smss}). It always depends on 
the initial state distribution  $\rho_{mm'}$ of Eq.~(\ref{inis}). In other words, the system cannot be thermalized 
with the reservoir \cite{Xiong2015,Xiong2020,Ali2020}.

\begin{widetext}

\begin{figure}[ht]
\includegraphics[width=15cm]{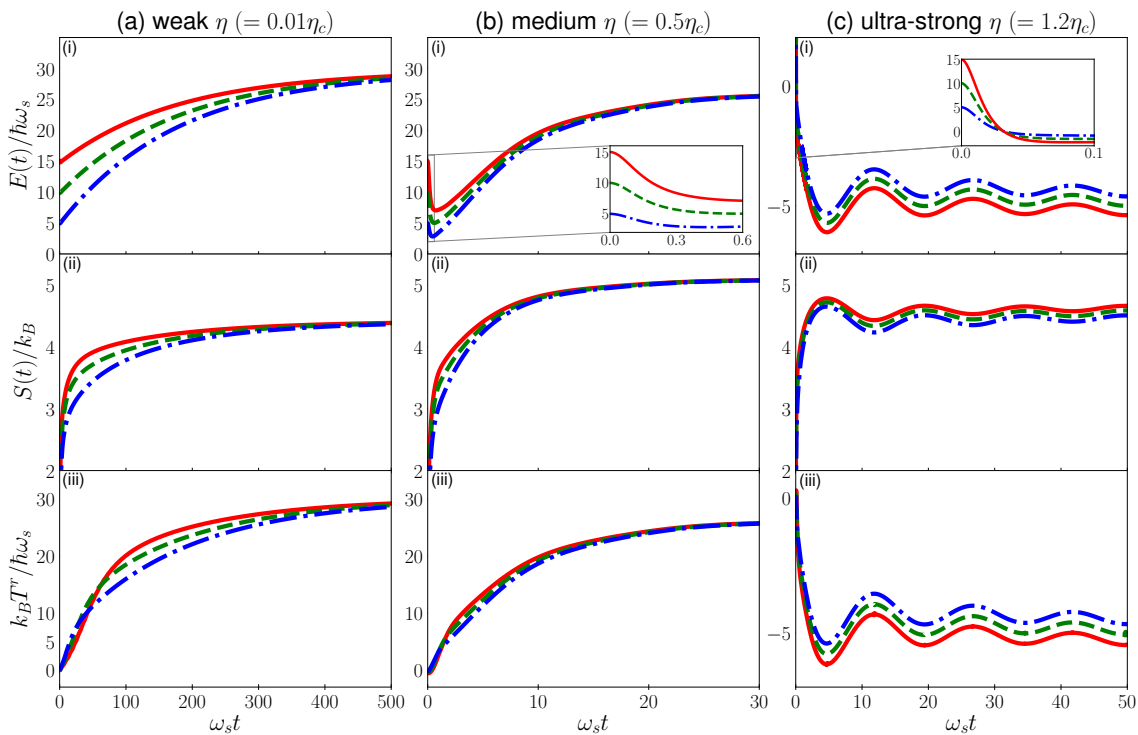}
\caption{\label{ultro_s}
The nonequilibrium dynamical evolution of the internal energy $U_\s(t)$, the entropy $S_\s(t)$ and the renormalized 
nonequilibrium temperature $T^r(t)$ for different system initial states $|n_0\rangle = |5\rangle, |10\rangle$ and $|15\rangle$ (corresponding 
to the blue dashed-dot line, the grees dashed line and the red solid line, respectively).  The left, the midden and the right 
panels correspond to  (a) the weak coupling ($\eta \ll \eta_c$), (b) the strong coupling $\eta < \eta_c$), and (c) the 
ultra-strong coupling $\eta > \eta_c$). The initial bath temperature $T_0=30 \hbar \omega_s$. Other parameters are taken 
the same as that in Fig.~\ref{pdsc}.}
\end{figure}

\end{widetext}

In Fig.~\ref{ultro_s}, we plot the nonequilibrium dynamical evolution of the internal energy, the entropy production and the 
renormalized temperature with different initial states for the very weak coupling ($\eta=0.01 \eta_c \ll \eta_c$), the strong 
coupling ($\eta =0.5 \eta_c < \eta_c$) and the ultra-strong coupling ($\eta=1.2 \eta_c > \eta_c$) cases. The results show 
that when $\eta > \eta_c$, different initial states of the system lead to different steady states. That is, the equilibrium hypothesis 
of the classical thermodynamics and statistical mechanics is broken down at ultra-strong coupling.  Also note that after consider 
the renormalization of the system Hamiltonian, the dynamics of the internal energy and the 
renormalized temperature are significantly changed, in particular in the strong coupling regime, in the comparison with our 
previous study \cite{Ali2020} where no renormalization of the system Hamiltonian is taken into account.  On the other hand,
regarding the system and the reservoir as a many-body system, the existence of the localized bound state in
the regime $\eta > \eta_c$ corresponds to a realization of the many-body localization \cite{Nandkishore15}. 
When the coupling strength $\eta$ crosses the critical value $\eta_c$, the transition from thermalization
to many-body localization occurs \cite{Nandkishore15}. Our exact solution provides the foundation of this transition 
between  thermodynamics and many-body localization.

\subsection{Quantum work and quantum heat}
Quantum mechanics does not introduce the concepts of work and heat because it deals with closed systems. 
For open quantum systems, 
the exchanges of matters, energies and information between the system and the reservoir cause the energy change 
of the system in the nonequilibrium evolution. This results in the work and chemical work (associated with chemical 
potential) done on the system or by the 
system, and the heat flowed into or out of the system. But usually the exchanges of matters, energies and information 
are correlated and interfered each other. This makes difficulties to define clearly the concepts of work, heat 
and chemical work in quantum thermodynamics. For the photons and phonons described by Eq.~(\ref{fH}),  no 
matters exchange between the system and the reservoir so that no chemical work involves (chemical potential is zero here).
Thus, the energy change of the system only involves with work and heat. After integrated out exactly the reservoir degrees 
of freedom, the reduced density matrix Eq.~(\ref{esrdm}) and the associated renormalized system Hamiltonian
Eq.~(\ref{rsH}) can be used to properly define thermodynamic work and heat within the quantum mechanics framework.
The chemical work will be considered when we study fermion systems, as we will discuss in the next section.
 
As it is shown from Eq.~(\ref{intE}), the nonequilibrium change of the internal energy in time contains two parts. 
One is the change (i.e.~the renormalization) of the system Hamiltonian $H^r_\s\!(t)$
(through the renormalization of the energy level $\hbar\w^r_\s(t)$) which corresponds to the \textit{quantum work} 
done on the system.  Note that in quantum mechanics, the concept of volume in a physical system is mainly characterized 
by energy levels through energy quantization. Thus, the change of volume is naturally replaced by the change of
energy levels, which results in a proper definition of work in quantum mechanics \cite{Zemansky1997}.
The other part is the change of the density state $\rho_\s(t)$ which corresponds to \textit{quantum heat}
associated with the entropy production. Consequently,
\begin{align}
\frac{dU_\s(t)}{dt} & =\!\Tr_\s\Big[\rho_\s(t)\frac{dH^r_\s\!(t)}{dt}\Big]+\Tr_\s\Big[H^r_{\s}\!(t)\frac{d\rho_\s(t)}{dt}\Big]  \notag \\
&= \frac{dW_s(t)}{dt}+\frac{dQ_s(t)}{dt} . 
 \label{1stl}
\end{align}
This is the first law of nonequilibrium quantum thermodynamics.
Thus, the quantum work and quantum heat can be naturally determined by
\begin{subequations}
\begin{align}
\frac{dW_\s(t)}{dt} &=\!\Tr_\s\Big[\rho_\s(t)\frac{dH^r_\s(t)}{dt}\Big]  = \overline{n}(t)\frac{d(\hbar\w^r_s(t))}{dt} , \\
\frac{dQ_\s(t)}{dt} & =\Tr_\s\Big[H^r_\s(t)\frac{d\rho_\s(t)}{dt}\Big]  =T^r\!(t) \frac{dS_\s(t)}{dt}, \label{heatt}
\end{align}
\end{subequations}
where $\overline{n}(t)= \Tr_\s[a^\+a\rho_\s(t)]$ is given by Eq.~(\ref{pnexp}).
The second equalities in the above equations have used explicitly the renormalized system Hamiltonian given after 
Eq.~(\ref{eme1}) and the definition of the renormalized temperature Eq.~(\ref{rdT}), respectively. 

In the literature, there are various definitions about work and heat for strong coupling quantum thermodynamics
but no consensus has been reached. The main concern is how to correctly include the system-reservoir coupling energy 
into the internal energy of the system \cite{Esposito2010b,Binder2015,Esposito2015,Alipour2016,Strasberg2017,
Perarnau2018,Rivas2020}.  The difficulty comes from the fact that most of open quantum systems cannot be solved
exactly so that it is not clear how to properly separate the contributions of the system-reservoir coupling interaction into
the system and into the reservoir, respectively. However, this difficulty can be overcome in our exact master equation 
formalism. Because we obtain the renormalized system Hamiltonian accompanied with the reduced density matrix after 
integrated out exactly all the reservoir degrees of freedom, namely exactly solved the partial trace over all the reservoir states. 
Thus,  the renormalized system Hamiltonian contains all possible 
contributions of the system-reservoir coupling interaction to the system energy. 

\begin{figure}[ht]
\centering
\includegraphics[width=8cm]{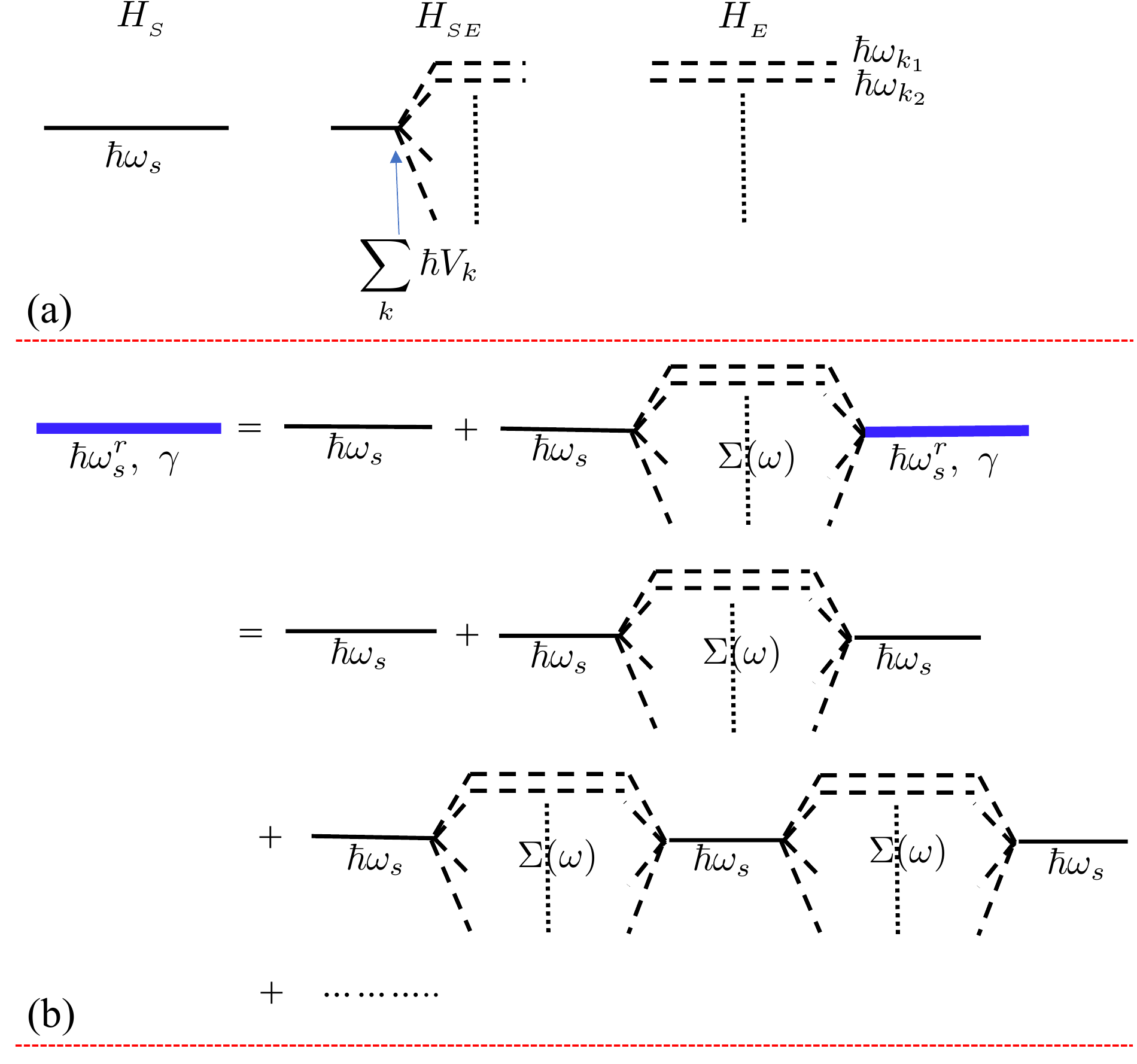}
\caption{(a) The diagrammatic spectra of the Hamiltonians of the system, the reservoir and the their interaction. 
(b) The diagrammatic Dyson expansion of Eq.~(\ref{ut}) in the energy domain, where 
$\Sigma(\omega)\!=\! \int d\omega' \frac{J(\omega')}{\omega-\omega'}$ is the self-energy arisen from the 
coupling between the system and the reservoir, and $J(\w)\equiv \sum_k|V_k|^2\delta(\w-\w_k)$. The renormalized 
system energy $\hbar\omega^r_s$ and the dissipation coefficient $\gamma$ of Eqs.~(\ref{re})-(\ref{ds}) are 
determined nonperturbatively from
Eq.~(\ref{ut}) with a Laplace transformation, which contains all order contributions upto the infinite orders from 
the system-reservoir coupling Hamiltonian, as shown in this diagrammatic expansion.}
\label{figrsf}
\end{figure}

Explicitly, let us rewrite the renormalized system Hamiltonian Eq.~(\ref{rsH}) as
\begin{align}
H_\s^r(t)& =\hbar \w^r_s(t,t_0)a^\+a = \hbar \w_sa^\+a + \delta\w_\s(t,t_0)a^\+a\notag \\
&=H_\s + \delta H_\s(t).   \label{rsh}
\end{align}
From Eq.~(\ref{re}) and Eq.~(\ref{ut}), we have 
\begin{align}
\w^r_s(t,t_0) &= \omega_\s \!+\delta \w_\s(t,t_0)  \notag \\
& =  \omega_\s \!+ \frac{1}{2}{\rm Im}\bigg[\! \int^t_{t_0} \!\! d\tau g(t,\tau) u(\tau,t_0)/ u(t,t_0)\bigg]. \label{rsf}
\end{align}
Here $H_\s= \hbar \w_sa^\+a $ is the bare Hamiltonian of the system. The second term in Eq.~(\ref{rsf})
contains all order contributions of the system-reservoir coupling interaction to the system energy, as shown in Fig.~\ref{figrsf}. Figure
\ref{figrsf}(a) is a diagrammatic plot of the bare Hamniltonians of the system, the reservoir and the interaction between them, respectively.  
Figure \ref{figrsf}(b) is the diagrammatic expansion (up to infinite orders) of the retarded Green function of Eq.~(\ref{ut}), from which all order 
renormalization effects to the system energy change (the system frequency shift) are reproduced. This diagrammatic expansion up to the infinite
orders illustrate the nonperturbative renormalized energy arisen from the system-reservoir interaction in our exact master equation theory. 

On the other hand, it is interesting to see that if we replace the full solution of the  Green function $u(\tau,t_0)$ approximately 
with the free-particle (zero-th order) Green function $u_0(\tau,t_0) = e^{-i\omega_\s (\tau-t_0)}$ (also
for $u(t,t_0)$) in Eq.~(\ref{rsf}),  the result is just the second-order renormalized energy correction.  By applying this same approximation 
to the dissipation and fluctuation coefficients in Eq.~(\ref{fmrc}), it is straightforward to obtain the time-dependent decay rate and 
noise in the Born-Markovian master equation, as we have shown in our previous work \cite{Xiong2010}. But once we have the exact master
equation with the exact solution, such approximated master equation is no longer needed.

\begin{figure}[ht]
\center
\includegraphics[width=5.5cm]{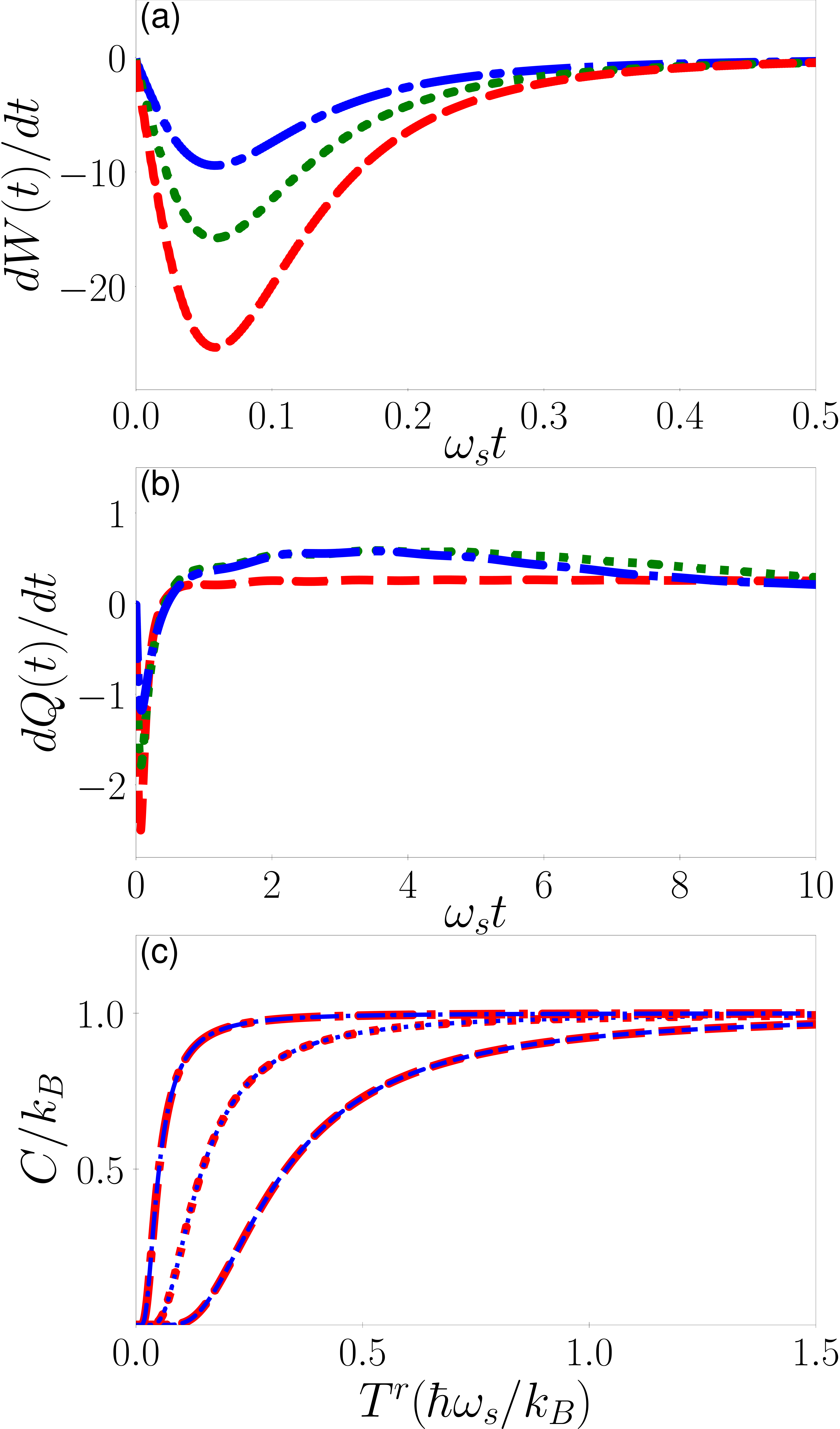}
\caption{\label{C} (a)-(b) The nonequilinrium evolution of quantum work and quantum heat changings with respect to the time,
$dW_\s(t)/dt$ and $dQ_\s(t)/dt$ (in the unit of $\hbar \omega_s^2$), for different coupling strengths.
(c) The steady-state specific heat as a function of the renormalized temperature calculated from the derivative of the internal energy
with respect to the renormalized temperature  
Eq.~(\ref{rhc}) (red lines) and from the partition function given in the Gibbs state Eq.~(\ref{rsspd}) for different 
initial temperatures. The dashed-dot, dot and dashed lines correspond to the different coupling strengths $\eta/\eta_c=0.3,0.5,0.8$, 
respectively. Other parameters are taken the same as that in Fig.~\ref{pdsc}.}
\end{figure}

In Fig.~\ref{C} (a)-(b), we plot the nonequilibrium evolution of $dW_\s(t)/dt$ and $dQ_\s(t)/dt$ for different coupling 
strengths. The negative values of $dW_\s(t)/dt$ show quantum work done by the system during the quantum mechanical time 
evolution, and more work is done by the system for the stronger system-reservoir coupling.  
While, $dQ_\s(t)/dt$ is negative and then becomes positive in time, which shows that quantum heat flows into the 
reservoir in the beginning and then flows back to the system in later time. 
This corresponds to the system dissipate energy very quickly into the reservoir in the very beginning, 
and then the thermal fluctuations arisen from the reservoir makes the heat flowing back slowly into the system. 
This heat flowing process can indeed be explained clearly from the exact master equation Eq.~(\ref{eme1}) 
combined with Eq.~(\ref{heatt}). It directly results in
\begin{align}
dQ_\s(t)/dt= \hbar \omega^r_s\big[-2\gamma(t,t_0)n(t) + \widetilde{\gamma}(t,t_0)\big],
\end{align}
where the first term is the contribution from dissipation and the second is the contribution of fluctuations in our
exact master equation. That is, the heating flow in open quantum systems is a combination effect of dissipation 
and fluctuation dynamics, which makes the system and the reservoir approach eventually to the equilibrium. This is
also a renormalization effect.

Furthermore, the quantum Helmholtz free energy of the system is defined by a Legendre transformation from 
the internal energy $U_\s(t)$  \cite{Ali2020a,Callen1985}:
\begin{align}
F_\s(t) & = U_\s(t) - T^r\!(t)S_\s(t) \notag \\ & \stackrel{t\ra \8}{\longrightarrow} -(1/\beta^r) \ln Z^r_\s .  \label{rfe}
\end{align}
From the above solution, we have 
\begin{align}
dF_\s(t) = dW_\s(t) - S_\s(t)dT^r\!(t),
\end{align} 
which naturally leads to the consistency that the quantum thermodynamic
work done on the system can be identified with the change of the Helmholtz free energy of
the system in isothermal processes \cite{Callen1985}. Moreover, the specific heat calculated from the
internal energy and from the Gibbs state with the renormalized Hamiltonian and temperature are also
identical, as shown in Fig.~\ref{C}(c),
\begin{align}
C =&\f{dQ_\s}{dT^r}= T^r \f{dS_\s}{dT^r} = \f{\pd U_\s}{\pd T^r}\bigg|_{\omega^r_s} ,  \label{rhc}
 \end{align}
 where the third thermodynamic law is justified from the specific heat at arbitrary coupling strength: $C \sim (T^r)^3$
 as $T^r\!\ra\!0$.  Thus, a consistent formalism of quantum thermodynamics from the weak coupling 
 to the strong coupling is obtained from a simple open quantum system of Eq.~(\ref{fH}).
 
\section{The more general formulation of quantum thermodynamics for all couplings}

\subsection{Multi-level open quantum system couple to multiple reservoirs}
The results from the exact solution of the single-mode bosonic  open system in the last section show that different from the previous investigations \cite{Seifert2016,Carrega2016,Ochoa2016,
Jarzynski2017,Marcantoni2017,Bruch2018,Perarnau2018,Hsiang2018,Anders2018,Strasberg2019,Newman2020,Ali2020a,Rivas2020},
only by introducing the renormalized temperature and incorporating with the renormalized system Hamiltonian,
can we obtain the consistent quantum thermodynamics for all coupling strengths.
Now we extend this quantum thermodynamics formulation to the more general situation: a multi-level 
system couples to multiple reservoirs (including both bosonic and fermionic systems) through
the particle exchange (tunneling) processes. 

In a quasiparticle picture, the Hamiltonian of a microscopic system in the energy eigenbasis
can be written as $H_S= \sum_i \varepsilon_{i} a^\+_ia_i$.  
As a specific example, consider the system be an individual system and the reservoir be 
a many-body system. The system Hamiltonian can be generally expressed as
\begin{align}
H_\s= \frac{{\bf P}^2}{2m} + V({\bf Q}) = \sum_i \varepsilon_i  |\psi_i\rangle \langle \psi_i |=
\sum_i \varepsilon_i a^\dag_i a_i .  \label{sph}
\end{align}
In Eq.~(\ref{sph}), the second equality is the spectral decomposition of the system Hamiltonian:
$H_\s |\psi_i\rangle= \varepsilon_i  |\psi_i\rangle$, and the last equality uses the second
quantization language: $ |\psi_i\rangle= a^\dag_i |0\rangle$ and $a_i |0\rangle=0$,
and $|0\rangle$ is the vacuum state.
The particle creation and annihilation operators $a^\dag_i$ and $a_i$ obey the
standard bosonic commutation and fermionic anticommutation relations: $[a_i, a^\dag_j]_\mp 
= a_ja^\dag_j \mp a^\dag_j a_i=\delta_{ij}$ when the system being boson and fermion systems, 
respectively. 

Similarly, the Hamiltonian of a reservoir can also be written as
$H_\e = \sum_{ k} \epsilon_{ k} b^\dag_{ k} b_{ k}$,
where $\epsilon_{ k}$ is usually a continuous spectrum and
could have band structure for structured reservoir. For a many-body reservoir in which 
the particle-particle interaction is not strong enough, the single quasiparticle picture works 
\cite{Thouless1972}. Then the reservoir Hamiltonian can be expressed approximated as
\begin{align}
H_\e &\simeq \! \sum_j \! \Big[\frac{\boldsymbol p^2_j}{2m_j} \!+\! U({\boldsymbol q}_j) 
\!+\! \overline{\sum_{j'}\!V({\boldsymbol q}_j, {\boldsymbol q}_{j'})} \, \Big]  \notag \\
& = \sum_k \epsilon_k |\psi_k\rangle \langle \psi_k| = \sum_k \epsilon_k b^\dag_k b_k,
\end{align}
where $\overline{\sum_{j'} V({\boldsymbol q}_j, {\boldsymbol q}_{j'})}$ represents 
the effective mean-field potential of many-body interactions, and $ \big[\frac{\boldsymbol p^2_j}{2m_j} 
+ U({\boldsymbol q}_j) + \overline{\sum_{j'}V({\boldsymbol q}_j, {\boldsymbol q}_{j'})}\,\big]|\psi_k\rangle = 
\epsilon_k |\psi_k\rangle$ gives the quasiparticle continuous spectrum of the reservoir. 
The reservoir particle creation and annihilation operators $b^\dag_k$ and $b_k$ also obey the
standard bosonic commutation or fermionic anticommutation relations. In fact,
the system can also be either a simple system or such a many-body system.

To dynamically address statistical mechanics and thermodynamics from quantum mechanical principle, the fundamental 
system-reservoir interactions are required to contain at least the basic physical processes of energy exchanges, matter 
exchanges and information exchanges between the system and reservoirs. 
The simplest realization for such a minimum requirement is the quantum tunneling Hamiltonian,
\begin{align}
H_{\s\e}= \sum_{ ik}\big(V_{ ik}a^\dag_i b_{ k}+ V^*_{ ik}b^\dag_{ k} a_i\big),
\end{align}
which is also the basic Hamiltonian in the study of quantum transport in mesoscopic physics as well as in nuclear, atomic 
and condensed matter physics for various phenomena \cite{Hang1996,Miroshnichenko2010,Jin2010,Lei2012}.
The coupling strengths $V_{ ik}$ are proportional to the quasiparticle wavefunction overlaps between 
the system and reservoirs and therefore are tunable through nanotechnology manipulations \cite{Hang1996,Miroshnichenko2010} 
so that they can be weak or strong coupling.  More discussions about fundamental system-reservoir interactions will be 
given in the next section.

Thus, a basic Hamiltonian with the minimum requirement for solving the foundation of quantum thermodynamics and statistical mechanics
can be modeled as 
\begin{align}
H_{\rm tot}(t) =&\, H_\s(t)+\sum_\alpha H^\alpha_\e(t)+\sum_\alpha H^\alpha_{\s\e}(t) \notag \\ 
=& \sum_i \varepsilon_{i}(t) a^\+_ia_i \!+ \!\! \sum_{\alpha k} \epsilon_{\alpha k}(t) b^\dag_{\alpha k} b_{\alpha k}  \notag \\
&+ \! \sum_{\alpha ik}\!\big[V_{\alpha ik}(t)a^\dag_i b_{\alpha k}\!+\! V^*_{\alpha ik}(t)b^\dag_{\alpha k} a_i\big], \label{qth}
\end{align}
which describes the system one concerned couples to multiple reservoirs. This is a generalization of the Fano-Anderson Hamiltonian 
we introduced \cite{Zhang2012,Zhang2018}. The index $\alpha$ stands for different reservoirs.
All parameters in the Hamiltonian can be time-dependently controlled with the current nano and quantum technologies.
This is an exact solvable Hamiltonian that involves explicit exchanges of energies, matter and information between the system 
and reservoirs. It allows us to rigorously solve quantum statistics and thermodynamics   
from the dynamical evolution of quantum systems. Also note that the above open quantum systems are different from the 
one proposed by Feynman and Vernon  \cite{Feynman1963} as well as by Caldeira and Leggett \cite{Leggett1983b} 
in the previous investigations of dissipative quantum dynamics in the sense that
their environment is made only by harmonic oscillators and the system-environment coupling is limited to 
the weak coupling. 

We have derived the  exact master equation of the open systems with Eq.~(\ref{qth}) for the reduced density 
matrix of the system. The formal solution of the total density matrix of the Liouville-von Neumann equation (\ref{voneq})
can be expressed as: 
\begin{align}  \label{fsLvNe}
\rho_\s(t)\!=\!\Tr_\e\big[{\cal U}(t,t_0)\big(\rho_\s(t_0) {\prod}_\alpha \!\!\!\X \rho^\alpha_\e(t_0)\big){\cal U}^\dag(t,t_0)\big],
\end{align}
where ${\cal U}(t,t_0)={\cal T}_{\ra}\exp\big\{\!-\!\frac{i}{\hbar}\!\int^t_{t_0}\!H_{\rm tot}(t')dt'\big\}$ is the time evolution 
of the total system, and ${\cal T_\ra}$ is the time-ordering operator. Here the system is initially in an arbitrary state $\rho_\s(t_0)$.
All reservoirs can be initially in their own 
equalibrium thermal states, $\rho^\alpha_\e(t_0)= e^{-\beta_{\alpha 0} (H^\alpha_\e - \mu_{\alpha 0}\hat{N}^\alpha)}/Z_\alpha$ 
which can have different initial temperature $\beta_{\alpha 0}=1/k_BT_{\alpha 0}$ and different chemical potentials $\mu_{\alpha 0}$ 
for different reservoir $\alpha$. Here $\hat{N}^\alpha$ is the total particle number operator of reservoir $\alpha$. 
After trace out all the environmental states through the coherent state path integrals \cite{Zhang1990}, the resulting exact master 
equation of the system is indeed a generalization of Eq.~(\ref{eme1}) to multi-level open systems
\cite{Tu2008,Jin2010,Lei2012,Yang2015,Yang2017,Zhang2018,Huang2020},
\begin{align}
\f{d}{dt}{\rho}_\s(t)   =& \f{1}{i\hbar}  \big[  H^r_\s(t),  \rho_\s(t)\big]  \!+\!\! {\sum}_{ij} \! \big\{\gamma_{ij}\left(  t,t_0\right)
\!\! \big[2a_{j}\rho_\s(t) a_{i}^{\+} \notag \\
&-\!a_{i}^{\+}a_{j}\rho_\s(t) \!-\!\rho_\s(t) a_{i}^{\+}a_{j}\big] \!+\!  \widetilde{\gamma}_{ij}(t,t_0) \big [a_{i}
^{\+}\rho_\s(t) a_{j} \notag \\
 &~~~{\pm} a_{j}\rho_\s(t)  a_{i}^{\+}\mp a_{i}^{\+}a_{j}
\rho_\s(t)  \!-\! \rho_\s(t) a_{j}a_{i}^{\+}\big]\big\}.
\label{EME}
\end{align}
where the upper and lower signs of $\pm$ correspond respectively to the bosonic and fermionic systems.

In the above exact master equation, all the renormalization effects arisen from the system-reservoir interactions have been taken 
into account when all the environmental degrees of freedoms are integrated out nonperturbatively and exactly in finding the 
reduced density matrix. These renormalization effects are manifested by the renormalized system Hamiltonian,
\begin{align}
H^r_\s(t)  ={\sum}_{ij} \varepsilon^r_{s,ij}(t,t_0)a_i^\+a_j  \label{fa_rH}
\end{align} 
and the dissipation and fluctuations coefficients  $\gamma_{ij}(t,t_0)$ and $\widetilde{\gamma}_{ij}(t,t_0)$ in Eq.~(\ref{EME}). 
These time-dependent coefficients are determined nonperturbatively and exactly by the following relations,
\begin{subequations}
\label{tddfc}
\begin{align}
& \varepsilon^r_{ij}(t,t_0) = \!- \hbar {\rm Im}\big[\dot{\bm u}(t,t_0)
\bm u^{-1}(t,t_0) \big]_{ij},   \label{fa_re} \\
& \gamma_{ij}(t,t_0) = \!-{\rm Re}\big[\dot{\bm u}(t,t_0)
\bm u^{-1}(t,t_0) \big]_{ij}, \\
& \widetilde{\gamma}_{ij}(t,t_0)   =\dot{\bm v}_{ij}(t,t)\!-\!\big[\dot{\bm u}(t,t_0) \bm u^{-1}(t,t_0)
\bm v(t,t)\!+\!\text{h.c.} \big]_{ij} .
\end{align}
\end{subequations} 
where ${\bm u}(t,t_0)$ and ${\bm v}(t,t)$ are $N\times N$ nonequilibrium Green function matrices and $N$ is the total number 
of energy levels in the system.

The nonequilibrium retarded Green functions $u_{ij}(t,t_0) \equiv \bra [a_i(t), a^\+_j(t_0)]_\pm$ which obeys the
equation of motion \cite{Zhang2012,Tu2008,Jin2010,Lei2012},
\begin{subequations}
\label{uvgf}
\begin{align}
\frac{d}{dt}\bm u(t,t_{0}) -& \f{1}{i\hbar} {\bm \varepsilon}(t) \bm u(t, t_{0}) +\!\!\int_{t_{0}}^{t}\!\! \!\! dt' \bm g(t,t')
\bm u(t',t_{0})  =0. \label{ute}
\end{align}
The nonequilibrium correlation Green function ${\bm v}(t,t)$ obeys the nonequilibrium  fluctuation-dissipation relation \cite{Zhang2012},
\begin{align}
{\bm v}(\tau,t) \!=\!\!\! \int^\tau_{t_0}\!\!dt_1\!\int^t_{t_0}\!\!dt_2\bm u(\tau,t_1)\,\widetilde{\bm g}(t_1, t_2)\,\bm u^\+(t,t_2).  \label{vt}
\end{align}
\end{subequations}
The integral memory kernels $\bm g(t,t')$ and $\widetilde{\bm g}(t_1, t_2)$ are
the system-reservoir time correlations and are given by  
\begin{subequations}
\label{giniti}
\begin{align}
& g_{ij}(t,t')  = \! \sum_{\alpha k} \f{1}{\hbar^2}V_{\alpha i k}(t')V^*_{\alpha jk}(t)
\exp\!\bigg\{\!\!-\!\f{i}{\hbar}\! \!\int_t^{t'} \!\!\!\! d\tau\epsilon_{\alpha k} (\tau)\bigg\} ,   \label{ik} \\
& \widetilde{g}_{ij}(t_1,t_2)  = \!\sum_{\alpha k}\f{1}{\hbar^2}V_{\alpha ik}(t_2)V^*_{\alpha jk} (t_1)
\big\langle b^\dag_{\alpha k}(t_0) b_{\alpha k}(t_0) \big\rangle_E  \notag \\
&~~~~~~~~~~~~~~~~~~~~\times
 \!\exp\!\bigg\{\!\!-\!\f{i}{\hbar}\! \!\int_{t_2}^{t_1} \!\!\!\!d\tau \epsilon_{\alpha k} (\tau)\bigg\} .
\end{align}
\end{subequations}
Here the initial reservoir correlation function,
\begin{align}
\big\langle b^\dag_{\alpha k}(t_0) b_{\alpha k}(t_0) \big\rangle_E & = f (  \epsilon_{\alpha k},T_{\alpha 0},\mu_{\alpha 0}) \notag \\
& =\frac{1}{[e^{(\epsilon -\mu_{\alpha 0})/k_{B}T_{\alpha 0}} {\mp} 1 ]}, 
\end{align} 
determines the initial particle distribution of the bosons or fermions in the initial thermal
reservoir $\alpha$ with the chemical potential $\mu_{\alpha 0}$ and the temperature 
$T_{\alpha 0}$ at initial time $t_0$. 
In the case the energy spectra of the reservoirs and the system-reservoir
couplings are time-independent, the memory kernels are simply reduced to 
$g_{ij}(t,t')\!=\!\!\!\int\!\!d\epsilon J_{ij}(\epsilon)e^{-i\epsilon(t-t')}$, 
$\widetilde{g}_{ij}(t_1,t_2)\! =\!\!\!\int\!\!d\epsilon J_{ij}(\epsilon)f(\epsilon,T_\alpha,\mu_\alpha)e^{-i\epsilon(t_1-t_2)}$,
where 
\begin{align}
J_{ij}(\epsilon) =\f{1}{\hbar^2} {\sum}_{\alpha k}V_{\alpha ik}V^*_{\alpha jk}\delta(\epsilon-\epsilon_{\alpha k})
={\sum}_\alpha J_{\alpha,ij}(\epsilon),
\end{align}
and $J_{\alpha ij}(\epsilon) $ is the spectral density matrix of reservoir $\alpha$.

\subsection{The theory of quantum thermodynamics from the weak to the strong couplings}
Again, if there exist no many-body localized bound states, the exact solution of Eq.~(\ref{EME}) has recently been solved 
\cite{Xiong2020} and its exact steady state is (see a detailed derivation in Appendix B)
\begin{align}
\rho^{\rm exact}_\s(t\!\ra\!\8)=\f{\exp\Big\{\sum_{ij}\Big(\!\ln\f{\overline{\bm n}}{I\pm\overline{\bm n}}\Big)_{ij}
a^\dag_ia_j\Big\}}{[\det(I\pm\overline{\bm n})]^{\pm 1}}
  \label{gss}
\end{align}
which is a generalized Gibbs-type state. Here 
$\overline{n}_{ij}=\lim_{t\ra \8} n_{ij}(t)$  is the one-particle density matrix defined  
as \cite{Jin2010,Yang1962}
\begin{align}
n_{ij}(t) \equiv \Tr_\s[a^\+_ia_j \rho_\s(t)] = \rho^{(1)}_{ij}(t).  \label{opdm}
\end{align} 
The solution Eq.~(\ref{gss}) remains the same for initial system-reservoir correlated states with a modification of 
$\widetilde{\bm g}(t_1,t_2)$ in Eq.~(\ref{vt}) to include the initial correlations between the system and 
reservoirs \cite{Yang2015,Huang2020}.
Thus, the nonequilibrium internal energy, entropy and particle number can be defined by
\begin{subequations}
\label{snd}
\begin{align}
&U_\s(t)\!\equiv\!\Tr_\s[H^r_\s\!(t)\rho_\s(t)] = \sum_{ij}\varepsilon^r_{ij}(t)n_{ij}(t),   \\
& S_\s(t)\equiv\!-\!k_B\Tr_\s[\rho_\s(t)\ln\rho_\s(t)], \\ 
& N_\s(t) \equiv\!\Tr_\s[a^\dag_i a_i \rho_\s(t)] \!=\!\ \sum_in_{ii}(t).  
\end{align}
\end{subequations}
They are related to each other and may form the fundamental equation for quantum thermodynamics
\cite{Callen1985,Ali2020a}: $U_\s(t) = U_\s(\varepsilon^r_s(t),S_\s(t), N_\s(t))$. Here energy levels
play a similar role as the volume \cite{Zemansky1997}. Thus,
\begin{align}
dU_\s(t)\!=\!dW_\s(t)+T^r(t)dS_\s(t)+\mu^r(t)dN_\s(t),
\end{align}
as the first law of nonequilibrium quantum thermodynamics.

Explicitly, the quantum work $dW_\s(t)$ done on the system is arisen from the changes of energy levels,
\begin{align}
\frac{dW_\s(t)}{dt}=\!\Tr_\s\Big[\rho_\s(t)\frac{dH^r_\s(t)}{dt}\Big] \!=\! {\sum}_{ij}n_{ij}(t)\frac{d\varepsilon^r_{s,ij}(t)}{dt}  .
\end{align}
The quantum heat $dQ_\s(t)$ (also including the chemical work $dW^c_\s(t)$)
comes from the changes of particle distributions and transitions (the one-particle density matrix, see Eq.~(\ref{opdm})),
\begin{align}
dQ_\s(t)+dW^c_\s(t)&
\!=\! {\sum}_{ij}\!\varepsilon^r_{s,ij}(t) dn_{ij}(t)\notag \\
&=\!T^r\!(t) dS_\s(t)\! +\! \mu^r\!(t)dN_\s(t).
\end{align}
It shows that $dn_{ij}(t)$ characterizes both the state information exchanges (entropy production)
and the matter exchanges (chemical process for massive particles) between the systems and
the reservoir. For photon or phonon systems, particle number is the number of
energy quanta $\hbar\w$ so that $\mu^r(t)\!=\!0$.
From the above formulation, we can define the renormalized temperature and renormalized chemical potential by
\begin{align}
&T^r\!(t)= \! \f{\pd U_\s(t)}{\pd S_\s(t)}\bigg|_{\varepsilon^r_s(t), N_\s(t)}, ~
\mu^r(t)= \! \f{\pd U_\s(t)}{\pd N_\s(t)}\bigg|_{\varepsilon^r_s(t), S_\s(t)} .   \label{rdTg}
\end{align}
  As a result, Eq.~(\ref{gss}) can be also written as the standard Gibbs state,
\begin{align}
\rho^{\rm exact}_\s(t\!\ra\!\8)=\f{1}{Z^r} \exp\big\{\!-\!\beta^r(H^r_\s\!-\!\mu^r \hat{N})\big\} , \label{ggss}
\end{align}
which is given in terms of the renormalized Hamiltonian $H^r_\s(t)$, the renormalized temperature $T^r(t)$ and the renormalized 
chemical potential $\mu^r(t)$ at steady state, and $\hat{N}=\sum_ia^\+_ia_i$ is the particle 
number operator of the system.
Because the exact solution of the steady state is a Gibbs state, thermodynamic
laws are all preserved at steady state. This completes our nonperturbative renormalization theory of quantum thermodynamics 
for all the coupling strengths.

\subsection{An application to a nanoelectronic system with two reservoirs}
\begin{figure}[ht]
\centering
\includegraphics[width=5.5cm]{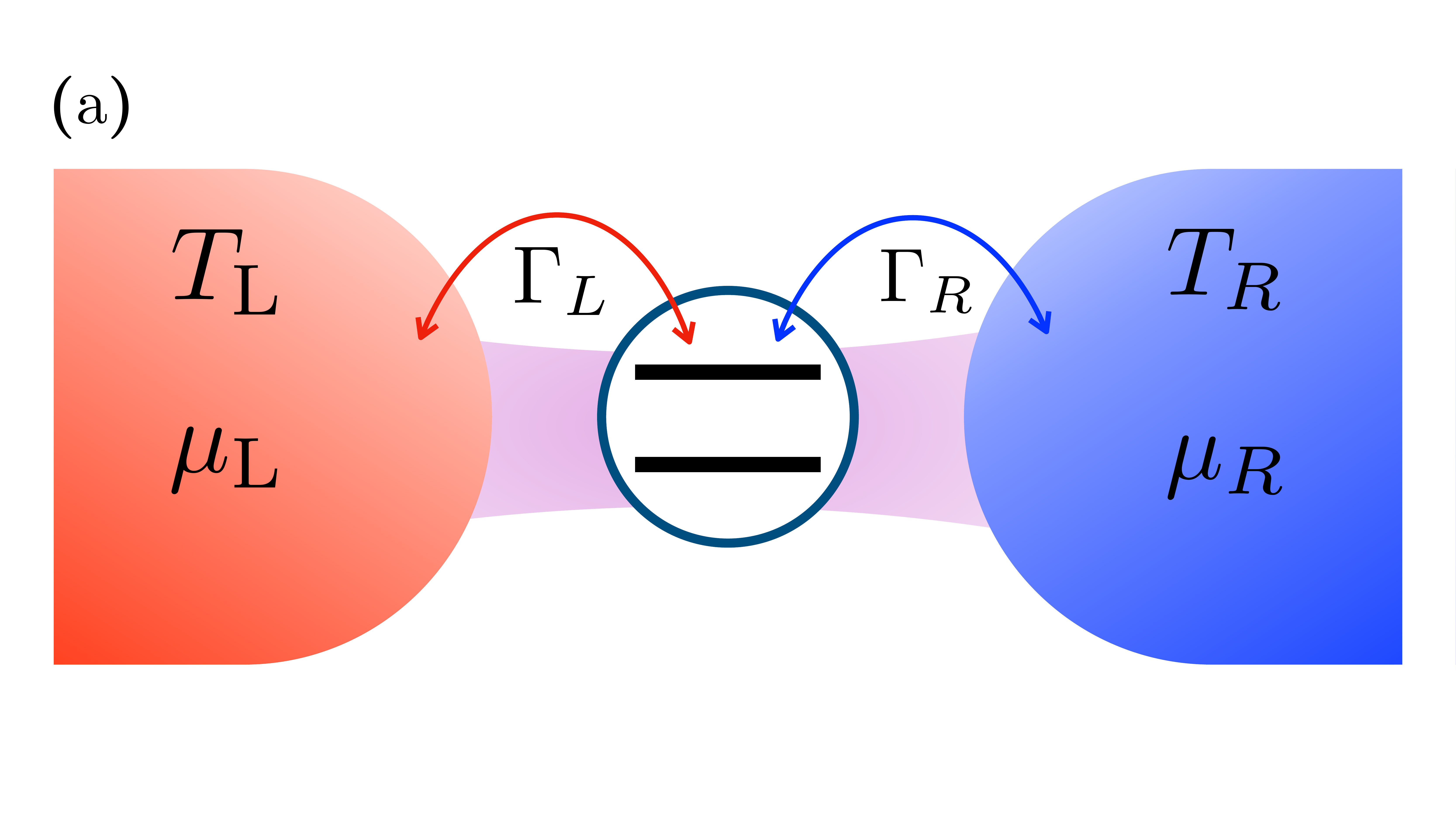}
\includegraphics[width=8cm]{Fig6b-g.pdf}
\includegraphics[width=8cm]{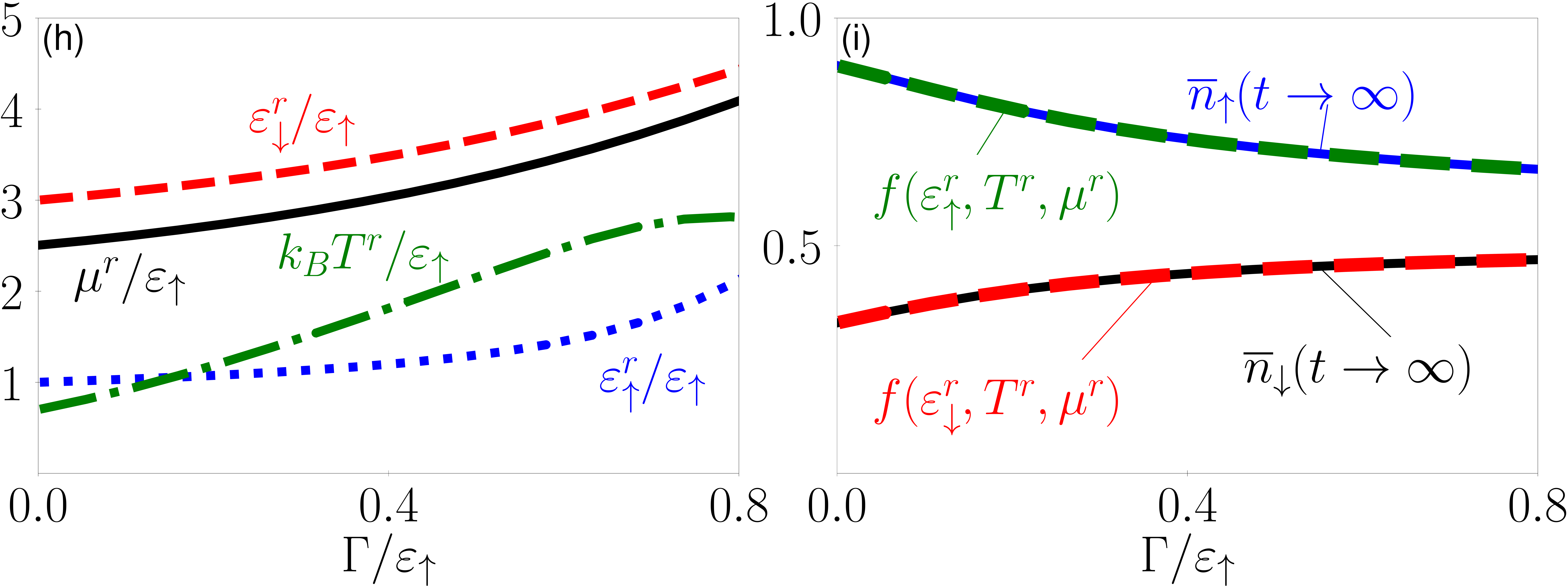}
\caption{(a) A schematic plot of the single-electron transistor device. (b)-(g) The nonequilibrium evolution of the energy levels 
$\varepsilon^r_{\uparrow,\downarrow}(t)$, the particle occupation in each level $\overline{n}_{\uparrow,\downarrow}(t)$, 
the internal energy $U_\s(t)$, the entropy $S_\s(t)$, the renormalized temperature $T^r(t)$ and chemical potential $\mu^r(t)$ 
at different coupling strength $\Gamma=0.2\varepsilon_\uparrow, 0.8\varepsilon_\uparrow$, respectively. (h) The steady-state 
value of the renormalized energy levels $\varepsilon^r_{\uparrow,\downarrow}$, the renormalized temperature $T^r$ and  the renormalized chemical
potential $\mu^r$ as a function of the coupling strength, and (i) the comparison of the renormalized Fermi-Dirac distribution 
$f(\varepsilon^r_{\uparrow,\downarrow},T^r,\mu^r)$ with
the exact solution of the $\overline{n}_{\uparrow,\downarrow}(t\!\ra\!\8)$ as a function of the coupling strength $\Gamma$.
Other parameters: $\varepsilon_\downarrow = 3 \varepsilon_\uparrow$, $k_BT_{L,R}=(3,0.1)\varepsilon_\uparrow$,
$\mu_{L,R}=(5,2)\varepsilon_\uparrow$, and $d=10\varepsilon_\uparrow$. 
}
\label{st}
\end{figure}

As a practical and nontrivial application:
we consider a nanoelectronic system,  the single electron transistor made of a quantum dot coupled
to a source and a drain. Here the two leads which are treated as two reservoirs \cite{Hang1996,Tu2008,Jin2010},
see Fig.~\ref{st}(a). The total Hamiltonian is 
\begin{align}
H_{\rm tot}\!=& \sum_{\sigma}
\varepsilon_\sigma a^\+_{\sigma} a_\sigma\!+\!\sum_{\alpha, k, \sigma}\epsilon_{\alpha  k}
c^\+_{\alpha k \sigma}c_{\alpha k \sigma} \notag \\
& +\!\sum_{\alpha, k, \sigma}(V_{\alpha k}a^\+_{\sigma}
c_{\alpha k \sigma}\!+\!V^*_{\alpha k}c^\+_{\alpha k \sigma} a_{\sigma}).  \label{setH}
\end{align}
The index $\sigma=\uparrow,\downarrow$ labels electron spin, $\alpha=L, R$ labels the left and right leads.
The two leads are setup initially in thermal states with different initial temperatures $T_{L,R}$ and chemical potentials
 $\mu_{L.R}$.  This is a prototype with nontrivial feature in the sense that two reservoirs initially 
have different temperatures and different chemical potentials so that when the system reaches the steady state, 
there exists only one final temperature and one final chemical potential. That is, one has to introduce 
the renormalized temperature $T^r$ and the renormalized chemical potential $\mu^r$ to characterize this final
equilibrium state when the system and two reservoirs reach
equilibrium.  While, other approaches proposed for strong coupling quantum thermodynamics in the 
last few years \cite{Seifert2016,Carrega2016,Ochoa2016,Jarzynski2017,
Marcantoni2017,Bruch2018,Perarnau2018,Hsiang2018,Anders2018,Strasberg2019,Newman2020,Rivas2020}
keep the reservoir temperature unchanged and therefore must be invalid for such simple 
but nontrivial open quantum systems. 

To explicitly solve the renormalized thermodynamics of the above system, let $|0\ket, |\uparrow\ket,|\downarrow\ket,|d\ket$ 
(the empty state, the spin up and down states
and the double occupied state, respectively) be the basis of the 4-dim dot Hilbert space of this quantum dot system.
Then the reduced density matrix has the form,
\begin{align}
\rho(t) = \matx{\rho_{00}(t) & 0 & 0 & 0 \\ 0 & \rho_{\uparrow\uparrow}(t) & \rho_{\uparrow\downarrow}(t) & 0 \\ 0 & 
\rho_{\downarrow\uparrow}(t) & \rho_{\downarrow\downarrow}(t) & 0 \\ 0 &0 & 0 & \rho_{dd}(t) } .
\end{align}
If the dot is initially empty, the $4\times4$ reduced density matrix has been solved exactly from the exact master equation  \cite{Tu2011,Yang2018}:
\begin{align}
&\rho_{00}(t)\!=\!\det[I\!-\!\bm v(t,t)], ~~ \rho_{dd}\!=\!\det[\bm {v}(t)], \notag \\
& \rho_{\uparrow\uparrow}(t)\!=\!v_{\uparrow\uparrow}(t)\!-\!\rho_{33}(t) , ~~ 
\rho_{\downarrow\downarrow}(t)\!=\!v_{\downarrow\downarrow}(t)\!-\!\rho_{33}(t), \notag \\
&\rho_{\uparrow\downarrow}(t)\!=\!v_{\uparrow\downarrow}(t)\!=\!\rho^*_{\downarrow\uparrow}(t).
\end{align} 
Here the $2\times 2$ matrix Green function $\bm{v}(t)\equiv \bm{v}(t,t)$ is determined 
by the Green function $\bm{u}(t,t')$. 
We take reservoir spectra as a Lorentzian form, then the spectral densities $J_\alpha(\epsilon)$ can be expressed as \cite{Meir1993,Tu2008}: 
\begin{align}
J_{\alpha,ij}(\epsilon)=\frac{\Gamma_\alpha d^2}{\epsilon^2+d^2}\delta_{ij}~~ (i,j=\uparrow,\downarrow),
\end{align} 
where $\Gamma_\alpha$ is the tunneling rate (the coupling strength) between the quantum dot and the lead $\alpha$. 
For simplicity, we also ignore the spin-flip tunneling. The exact solution of the reduced density matrix is rather simple, 
\begin{align}
\rho(t)  = \det[1-\bm{v}(t)]\exp \Big\{ \bm{a}^\dag \ln\frac{\bm{v}(t)}{1-\bm{v}(t)} \bm{a}\Big\}.
\end{align}
Here $\bm{a}^\dag=(a^\dag_\uparrow, a^\dag_\downarrow)$, and
\begin{subequations}
\label{setvt}
\begin{align}
&v_{ii}(t) = \!\!\int^t_{t_0}\!\!\!dt_1\!\!\int^t_{t_0}\!\!\!\!dt_2\,u_{ii}(t,t_1)\,\widetilde{g}(t_1, t_2)\,u^*_{ii}(t,t_2),  \\
&v_{\uparrow \downarrow}=0 , \\
& \frac{d}{dt}u_{ii}(t,t_0) + i\varepsilon_{i} u_{ii}(t,t_0) 
+ \!\!\int_{t_0}^t\!\!\! dt' \!\! \int \!\! d\epsilon\frac{\Gamma d^2 e^{-\epsilon(t-t')}}{\epsilon^2+d^2}u_{ii}(t,t') \notag \\
&~~~~~~~~~~~~~~~~~~~~~~~~~~~~~~~~~~~~~~~~~~~~~~~~=0 ,
\end{align}
\end{subequations}
for $i=\uparrow, \downarrow$, and $\Gamma=\Gamma_{L}+\Gamma_{R}$.
As a result, the nonequilibrium internal energy, the entropy and the total average particle number can be found analytically,
\begin{subequations}
\begin{align}
 U_\s(t)= & \varepsilon^r_\uparrow(t) v_{\uparrow\uparrow}(t) + \varepsilon^r_\downarrow(t) v_{\downarrow\downarrow}(t) ,  \\
 S_\s(t)= & -v_{\uparrow\uparrow}(t)\ln v_{\uparrow\uparrow}(t)-v_{\downarrow\downarrow}(t)\ln v_{\downarrow\downarrow}(t) \notag \\
 & -(1-v_{\uparrow\uparrow}(t))\ln (1-v_{\uparrow\uparrow}(t)) \notag \\
 & -(1-v_{\downarrow\downarrow}(t))\ln (1-v_{\downarrow\downarrow}(t)), \\
N_\s(t) = & v_{\uparrow\uparrow}(t) + v_{\downarrow\downarrow}(t) .   \label{setN}
\end{align}
\end{subequations}
From the above solution, the corresponding renormalized energy, renormalized temperature and renormalized chemical potential 
can be calculated straightforwardly.

In Fig.~\ref{st}(b)-(g), we show the nonequilibrium evolution of the renormalized energy levels $\varepsilon^r_{\uparrow,\downarrow}(t)$, 
the particle occupations in each levels $\overline{n}_{\uparrow,\downarrow}(t)$, the internal energy $U_\s(t)$, the entropy $S_\s(t)$, 
the renormalized temperature $T^r(t)$ and the renormalized chemical potential $\mu^r(t)$ for the coupling 
strength $\Gamma_L =\Gamma_R=\Gamma/2$ for different coupling strengths. It shows that in such a nano-scale device, 
all physical quantities are quickly approach to the steady state. Then, in Fig.~\ref{st}(h), we plot the steady-state values of the 
renormalized energy levels $\varepsilon^r_{\uparrow,\downarrow}$, the renormalized temperature $T^r$ and  the renormalized chemical
potential $\mu^r$ as a function of the coupling strength, respectively. These renormalized thermodynamical quantities 
change as the change of the coupling strength.
Finally, in Fig.~\ref{st}(i), we present the corresponding renormalized Fermi-Dirac distributions (the Fermi-Dirac distribution 
with renormalized energy, the renormalized temperature and the renormalized chemical potential):
$f(\varepsilon^r_{\uparrow,\downarrow},T^r,\mu^r)=1/[e^{(\varepsilon^r_{\uparrow,\downarrow}-\mu^r_\alpha)/k_{B}T^r_\alpha}+1]$.
We compare the  renormalized Fermi-Dirac distributions  with the exact solution of the occupation numbers  
$\overline{n}_{\uparrow,\downarrow}(t\!\ra\!\8)$
which are solved from the exact master equation. The results shows that they completely agree with each other.
This provides the proof to the consistency of the renormalized strong coupling quantum
thermodynamics for fermionic systems. 

\begin{figure}[ht]
\centering
\includegraphics[width=8.5cm]{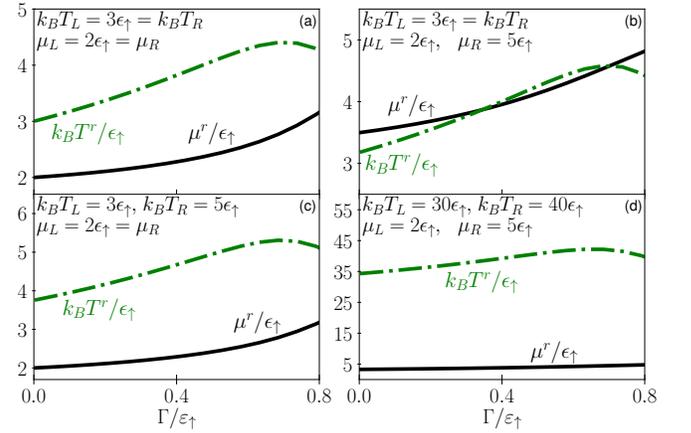}
\caption{The steady-state renormalized temperature and  renormalized chemical potential of the single-electron transistor as a function
of the system-reservoir coupling strength $\Gamma$ changing from weak to strong for different setups of the initial temperatures and initial 
chemical potentials of the two leads (reservoirs): (a) Two reservoirs have the same initial temperature and chemical potential; (b) The initial 
temperatures of the two reservoirs are the same but their initial chemical potentials are different; (c) The initial chemical potentials 
of the two reservoirs are the same but their initial temperatures are different; and (d) Two reservoirs at high temperature limit.
}
\label{dist}
\end{figure}

To have a clearer physical picture about the renormalized temperature and  renormalized chemical potential when the system 
coupled to two reservoirs, we take various different setups of the initial temperatures and initial chemical potentials of the two reservoirs 
in Fig.~\ref{dist}. From these results, we can see how the renormalized temperature and the renormalized chemical potential changes 
for the different setups, even in the weak-coupling regime. To understand these results, we first compare the exact solution with its 
weak-coupling limit. Since we also take the same spectral density for two reservoirs, we find that in the very weak-coupling limit (WCL),
\begin{align}
N_\s(t\rightarrow \infty)  & =  n_{\uparrow}(\mu^r,T^r) + n_{\downarrow}(\mu^r,T^r) \notag \\
& \stackrel{\rm WCL}{\rightarrow}  \frac{1}{2}\big[n_{\uparrow}(\mu_L,T_L) + n_{\uparrow}(\mu_R,T_R) \big] \notag \\
&~~~~~~~ +\frac{1}{2}\big[n_{\downarrow}(\mu_L,T_L) + n_{\downarrow}(\mu_R,T_R)\big] .  \label{setNwcl}
\end{align}
The first equality is the exact solution from Eqs.~(\ref{setN}) and the second equality is obtained with
the help (\ref{setvt}) in the very weak-coupling limit, where $\mu_L,T_L$ and $\mu_R,T_R$ are the initial chemical 
potential and temperatures of the left and right reservoirs, respectively. 
Figure \ref{dist}(a) shows the results for the two reservoirs that have the same initial 
temperature and the same initial chemical potential. Because the two reservoirs are set to have the same spectral density,
the two reservoirs are equivalent to one single reservoir in this case. Thus, the renormalized temperature and the 
chemical potential approach to the initial temperature and the initial 
chemical potential in the very weak coupling limit, as shown in Fig.~\ref{dist}(a), also as we expected.
Figure \ref{dist}(b) shows the results for the two reservoirs sharing the same initial temperature but having different initial chemical potentials. 
Naively, one may think that the renormalized temperature in the very weak coupling limit should be the same as the same initial temperature of 
the two reservoirs and the renormalized chemical potential should be $\mu^r=(\mu_L +\mu_R)/2$. From Fig.~\ref{dist}(b), we see that
the renormalized chemical potential in the very weak coupling limit is $\mu^r=(\mu_L +\mu_R)/2$, as we expected from energy 
conservation law. However, the renormalized temperature is a bit larger than the initial temperature. This result can actually 
be understood from Eq.~(\ref{setNwcl}).  Because $\mu^r=(\mu_L+\mu_R)/2 \neq \mu_L\neq \mu_R$,  Eq.~(\ref{setNwcl}) shows 
that $T_L\neq T^r\neq T_R$  in the very weak-coupling limit, even through $T_L=T_R$. 
Figure \ref{dist}(c) shows further the case $\mu_L=\mu_R$ and $T_L \neq T_R$. We have $\mu^r=\mu_L=\mu_R$, and from 
Eq.~(\ref{setNwcl}), we find that $T^r \neq (T_L+T_R)/2$ in the very weak coupling limit, as shown in Fig.~\ref{dist}(c). 
Figure \ref{dist}(d) shows the high temperature limit in which the chemical potentials play a little role. Thus, we have 
$T^r\simeq (T_L+T_R)/2$ in the very weak coupling limit, even if $\mu_L \neq \mu_R$. This is shown in  
Fig.~\ref{dist}(d). These results demonstrate that only at 
very high temperature, the renormalized temperature  $T^r=(T_L+T_R)/2$.  In other words, in the 
quantum regime, the renormalized temperature we introduced is necessary even at very weak-coupling limit for multi-reservoirs. 
This justifies further the consistency of our renormalized theory for quantum thermodynamics at any coupling.

\section{Extension to more arbitrary system-reservoir couplings}
The renormalized quantum thermodynamics for arbitrary coupling strength presented in Sec.~III,
given by Eq.~(\ref{snd}) to Eq.~(\ref{ggss}), is formulated from the exact master equation (\ref{EME}) based on the 
system-reservoir coupling of Eq.~(\ref{qth}). However, this formulation can be directly extended to general open quantum 
systems with system-reservoir couplings not limiting to the form of Eq.~(\ref{qth}).  This is because
the renormalized Hamiltonian $H^r_\s(t)$ is determined by the nonequilibrium Green function $\bm u(t,t_0)$ 
of Eq.~(\ref{ute}). It can be applied to arbitrary system interacting with arbitrary environment. In Eq.~(\ref{ute}), $\bm g(t,t')$ 
is the self-energy correlation that can be easily generalized to any interacting system using the nonequilibrium Green 
function technique in many-body systems \cite{Kadanoff1962,Keldysh1965}. 
Meanwhile, the renormalized temperature $T^r(t)$ (the renormalized chemical potential
$\mu^r(t)$) of Eq.~(\ref{rdTg}) are determined by the changes of internal energy of the system with respect to the changes of  
the von Neumann entropy (the average particle number of the system). These nonequilibrium thermodynamic quantities are 
well defined by Eq.~(\ref{snd}). 
They rely neither on the exact master equation of Eq.~(\ref{EME}) nor the system-environment coupling in
the Hamiltonian of Eq.~(\ref{qth}). They are all determined by the reduced density matrix which can be solved from
the Liouville-von Neumann equation (\ref{voneq}), whose formal solution can be expressed as
\begin{align}
\rho_\s(t) = \Tr_\e[{\cal U}(t,t_0) \rho_{\rm tot} (t_0){\cal U}^\+(t,t_0)] .   \label{fs_rho2}
\end{align}
Here, ${\cal U}(t,t_0)$ is the quantum evolution operator, the same as the one given after 
Eq.~(\ref{fsLvNe}) but the total Hamiltonian can be extended to arbitrary system interacting with arbitrary reservoir. 
In most of cases, taking the trace over the environmental states  
is the most difficult problem in open quantum systems. Practically, one can use the 
perturbation expansion method to calculate the trace over the environmental states order by order approximately
\cite{Breuer2008}, or using the coherent state path integrals to nonperturbatively trace over all the 
environmental states  as we did \cite{Tu2008,Jin2010,Lei2012,Yang2015,Zhang2018,Huang2020,Zhang1990,Zhang2022}.  
Here we focus on the nonperturbation procedure. All the renormalization effects of the system-reservoir interactions
on the system can be obtained from this procedure.

To be specific, let us consider a general fermionic system coupled to a general bosonic reservoir. Notice that Eq.~(\ref{qth})
describes the exchanges of energies and particles between the system and the reservoir only for both the system and the 
reservoir that are made of the same type of quasiparticles, either bosons or fermions. When the system is a fermionic system 
and the reservoir is a bosonic system, the system-reservoir coupling Hamiltonian generally has the following interaction form \cite{Zhang2012,Zhang2018}
\begin{align}
H_{\s\e}=\sum_{ ijk}\big[V_{ij}(k)c^\dag_i c_jb^\dag_k + V^*_{ij}(k) c^\dag_j c_i b_k \big].   \label{epi}
\end{align}
This system-reservoir interaction describes the energy exchange between the system and the reservoir through the transition 
of a fermion (e.g.~an electron) between two states by emitting a boson (a quanta 
 of energy, such as a photon or a phonon) into the reservoir or absorbing a boson from the reservoir. The creation and 
annihilation operators $c^\dag_i, c_i$~($b^\dag_k, b_k$) obey the standard fertmionic anticommutation (bosonic commutation) 
relationships. In fact, Eq.~(\ref{epi}) is the general form of the non-relativistic electron-photon interaction that can be derived from the fundamental 
field theory of quantum electrodynamics (QED). 

Explicitly, the QED Lagrangian determines the fundamental electron-photon interaction as follows  \cite{Peskin1995},
\begin{align}
{\cal L}_{\rm QED} = \overline{\psi}(i\gamma^\mu \partial_\mu - m)\psi - \frac{1}{4}F^{\mu\nu}F_{\mu\nu} 
- e\overline{\psi}\gamma^\mu A_\mu \psi, \label{lqed}
\end{align}
where $\psi(x)$ is the fermionic field for electrons, $A_\mu(x)$ is the covariant 4-vector of the electromagnetic (EM) field, $\gamma^\mu$ 
is the Dirac matrix, and $F_{\mu\nu}(x)=\partial_\mu A_\nu(x) - \partial_\nu A_\mu(x)$ is the EM field strength tensor. 
The first two terms of Eq.~(\ref{lqed}) leads to the free electron and photon Hamiltonians. The last term gives the fundamental 
electron-photon interaction in the non-relativistic limit (by ignoring positrons and choosing a proper gauge). 
Thus, the non-relativistic QED Hamiltonian is given by \cite{Coulomb_gauge}  
\begin{align}
\!\!\!H_{\rm QED} \!= & H_{\rm electron} \!+\! H_{\rm photon} \!+\! H_{\rm e-p}  \notag  \\
= &\!\sum_{\bm p}\! \varepsilon_{\bm p} c^\dag_{\bm p} c_{\bm p} 
\!+\!\!\!\! \sum_{\bm p, \bm p', \bm q} \!\!\! U(\bm q) c^\dag_{\bm p+\bm q}c^\dag_{\bm p'\!-\!\bm q} c_{\bm p'} c_{\bm p}
\!+ \!\! \sum_{\bm k} \! \hbar \omega_{\bm k} b^\dag_{\bm k} b_{\bm k}  \notag \\
& +\! \hbar\!\sum_{\bm p \bm k} \!\big[V({\bm k}) c^\dag_{\bm p} c_{\bm p-\bm k} b_{\bm k} \!+\! V^*({\bm k}) 
c^\dag_{\bm p- \bm k} c_{\bm p } b^\dag_{\bm k} \big],   \label{hqed}
\end{align}
where $c^\dag_{\bm p }, c_{\bm p}$ and $b^\dag_{\bm k},b_{\bm k}$ are creation and annihilation operators of electrons  and 
photons with momentum $\bm{p}$ and $\bm{k}$. The summations over $\bm{p}$ and $\bm{k}$ should be replaced by the 
continuous integrals $\int\!\!\frac{d^3\bm p}{(2\pi)^3}$ and $\int\!\!\frac{d^3\bm k}{(2\pi)^3}$. Also, without loss of generality, 
we have omitted the indices of electron spin and photon polarization. The first term in the second equality in Eq.~(\ref{hqed}) is the free electron
Hamiltonian. The second term is the electron-electron Coulomb interaction arisen from the choice of Coulomb gauge. The
third term is the EM field Hamiltonian, and the last two terms are the electron-photon interaction. Note that the electron-phonon 
interaction in solid-state physics has the same form.
Equation (\ref{hqed}) can describe most of non-relativistic physics in the current physics research, unless one is also interested in the
phenomena in the smaller scale of nuclear arisen from the weak and strong interactions or the larger scale of universe from gravity.

In the following, we shall find all the nonperturbative renormalization effects on electrons from the electron-photon interaction 
by using the coherent state path integrals \cite{Zhang1990} to nonperturbatively trace over all the environmental states. 
To do so, we may express the exact reduced density $\rho_\s (t) $  of Eq.~(\ref{fs_rho2}) as
$\rho_\s (\bm{\xi}^\dag_f, \bm{\xi}'_f, t )   = \langle \bm{\xi}_f |\rho_\s(t) | \bm{\xi}'_f \rangle$ which
is generally given by
\begin{align}
\rho_\s (\bm{\xi}^\dag_f, \bm{\xi}'_f, t )    
=  \!\! \int   \!\! d\mu & (  \bm{\xi}_0)  d\mu (\bm{\xi}'_0 ) \rho_\s( \bm{\xi}_0,\bm{\xi}'_0,t_0)  \notag \\ 
 & ~~~\times \! \mathcal{J}_{\rm QED}\!( \bm{\xi}^\dag_f,\bm{\xi}'_f, t; \bm{\xi}_0, \bm{\xi}'^\dag_0,t_0) . \label{crhos_t}
\end{align}
Here we have utilized the unnormalized fermion coherent states $|\bm{\xi}\rangle \equiv \exp(\sum_{\bm p} \xi_{\bm p} 
c^{\dag}_{\bm p})|0\rangle$. The integral measure $d\mu ( \bm{\xi})  = \prod_{\bm p} d\xi^*_{\bm p} d\xi_{\bm p} 
e^{- |\xi_{\bm p}|^2}$ is defined the Haar measure in Grassmannian space. The vectors $ \bm{\xi} \equiv (\xi_{\bm{p}_1}, 
\xi_{\bm{p}_2}, . . .)$ is a one-column matrix and $\xi^*_{\bm p_i}$ is a Grassmannian variable. 
The  propagating function $\mathcal{J}\!(\bm{\xi}^\dag_f, \bm{\xi}'_f, t; \bm{\xi}_0, {\bm \xi'}_0^\dag, t_0)$
in Eq.~(\ref{crhos_t}), which describes the nonequilibrium evolution of the states of the electron system from the initial state
$\rho_\s(t_0)$ to the state at any later time $\rho_\s(t)$, can be obtained analytically after completing exactly the path 
integrals over all the photon modes. 
The result is %
\begin{align}
\mathcal{J}(\bm{\xi}^\dag_{t}, \bm {\xi}'_{t}, t; \, & \bm \xi_0,{\bm \xi'_0}^{\dag},  t_0)= \int\mathcal{D}[ \bm \xi;  \bm \xi']\exp\Big\{\frac{i}{\hbar}
\big(S_{s}[\bm \xi^\dag,\bm \xi] \notag \\
&-S^*_{s}[\bm \xi'^\dag, \bm \xi'] +S^{\rm qed}_{\rm IF} [\bm \xi^\dag,\bm \xi; \bm \xi'^\dag, \bm \xi'] \big)\Big\} ,
\label{ppg}
\end{align}
where $\mathcal{D}[ \bm \xi;  \bm \xi']= \prod_{\bm p,t<\tau<t_0} d\xi^*_{\bm p}(\tau)d\xi_{\bm p}(\tau)$, 
$S_{e}[\bm \xi^\dag,\bm \xi]$ is the bare electron action of the original free-electron Hamiltonian plus the electron-electron 
Coulomb interaction in QED. In the fermion coherent state representation, it is given by
\begin{align}
S_{e}[\bm \xi^\dag,\bm \xi] = & -{i\hbar \over 2}\big[\bm \xi^\dag_t \bm{\xi}(t) + \bm \xi^\dag(t_0) \bm{\xi}_{t_0} \big]  \notag \\
&+\!\! \int_{t_0}^{t}\!\! d\tau \Big\{ {i\hbar \over 2}
\big[\dot{\bm \xi}^\dag(\tau) \bm{\xi}(\tau)  -\bm \xi^\dag(\tau) \dot{\bm{\xi}}(\tau)\big] \notag \\
& ~~~~~~~~~~~~~~~~~- {\cal H}(\bm \xi^\dag(\tau), \bm{\xi}(\tau)) \Big\},  \label{qed_saction}
\end{align}
where ${\cal H}(\bm \xi^\dag(\tau), \bm{\xi}(\tau))= \sum_{\bm p} \varepsilon_{\bm p} \xi^*_{\bm p}(\tau) \xi_{\bm p}(\tau) 
+ \sum_{\bm p, \bm p', \bm q} U(\bm q) \xi^*_{\bm p+\bm q}(\tau) \xi^*_{\bm p'\!-\!\bm q}(\tau) \xi_{\bm p'}(\tau) \xi_{\bm p}(\tau)$.  
The additional action $S^{\rm qed}_{\rm IF} [\bm \xi^\dag,\bm \xi; \bm \xi'^\dag, \bm \xi']$ 
in Eq.~(\ref{ppg}) is an electron action correction arisen from electron-photon interaction after exactly integrated out all the photon 
modes \cite{Zhang2022}:
\begin{widetext}
\begin{align}
S^{\rm qed}_{\rm IF} [\bm \xi^\dag,\bm \xi; \bm \xi'^\dag, \bm \xi']=
\!\!  \int_{t_0}^{t}\!\!\! & d\tau \Bigg\{ i\hbar  \sum_{\bm{p}\bm{p}'\bm{k}}  \bigg[ \int_{t_0}^{\tau} \!\!\! d\tau' \!
\Big\{ \sigma^+_{\bm p,\bm k}(\tau) G_{\bm k} (\tau,\tau') \sigma^-_{\bm p',\bm k}(\tau')
+ \sigma'^-_{\bm p',\bm k}(\tau) G^*_{\bm k} (\tau,\tau') \sigma'^+_{\bm p,\bm k}(\tau') \!\Big\} \notag  \\
&- \!\!\! \int_{t_0}^{t} \!\!\! d\tau' \Big\{ \sigma'^+_{\bm p,\bm k}(\tau) G_{\bm k} (\tau,\tau') \sigma^-_{\bm p',\bm k}(\tau')  
\!-\! \big[\sigma^+_{\bm p,\bm k}(\tau) \!-\! \sigma'^+_{\bm p,\bm k}(\tau)\big]  \widetilde{G}_{\bm k}(\tau,\tau') 
\big[\sigma^-_{\bm p',\bm k}(\tau') \!-\! \sigma'^-_{\bm p',\bm k}(\tau')\big] \!\Big\} \bigg] \! \Bigg\} \label{qed_ation}
\end{align}
\end{widetext}
This is a generalization of the Feynman-Vernon influence functional \cite{Feynman1963} to electron-photon interacting systems
so that we may also call the action of Eq.~(\ref{qed_ation}) as the influence functional action.
For simplicity, here we have introduced the composite-particle variables $\sigma^+_{\bm p,\bm k}(\tau) \equiv 
\xi^*_{\bm p}(\tau) \xi_{\bm p-\bm k}(\tau)$
and $\sigma^-_{\bm p,\bm k}(\tau) \equiv \xi^*_{\bm p-\bm k}(\tau) \xi_{\bm p}(\tau)=(\sigma^+_{\bm p,\bm k}(\tau))^\dag$, 
which correspond  to the spin-like variables of the exciton operators $a^\dag_{\bm p }a_{\bm p- \bm k}$ and 
$a^\dag_{\bm p - \bm k}a_{\bm p}$, respectively.  The non-local time correlations in Eq.~(\ref{qed_ation}) are given by
\begin{subequations}
\label{ggbeta}
 \begin{align}
  \label{gvt}
 G_{\bm k}(\tau,\tau') &=|V(\bm{k})|^2 e^{-i \omega_{\bm k}(\tau-\tau')},
 \\
 \widetilde{G}_{\bm k}(\tau,\tau') &=|V(\bm{k})|^2 \overline{n}(\omega_{\bm k}, T_0)  e^{-i \omega_{\bm k}(\tau-\tau')}
 \label{gbetavt}
 \end{align}
 \end{subequations}
which depict the time-correlations between electrons and photons. The four terms in 
Eq.~(\ref{qed_ation}) come from the contributions of the electron-photon interactions to the electron forward propagation, the electron backward 
propagation, and to the electrons mixed from the forward with backward propagations at the end point time $t$ and at the initial time $t_0$, 
respectively, through the path integrals over all the photon modes.

The above results show that the propagating function Eq.~(\ref{ppg}) of the reduced density 
matrix for electrons in non-relativistic QED and the corresponding influence functional action Eq.~(\ref{qed_ation}) 
have the same form as that for our generalized Fano-Anderson Hamiltonian Eq.~(\ref{qth}), as shown by 
Eqs.~(\ref{ppg1}) and (\ref{fa_ation}) in Appendix B. The main difference is that the bare system action Eq.~(\ref{saction}) 
and the influence functional action Eq.~(\ref{fa_ation}) for the generalized Fano-Anderson Hamiltonian are quadratic 
with respect to the integrated variables in the path integrals of the propagating function for the reduced density matrix. 
They can be solved exactly and the resulting propagating 
function  is given by Eq.~(\ref{rdmcr1}) in Appendix B. Here the bare electron action Eq.~(\ref{qed_saction}) and the influence functional
action Eq.~(\ref{qed_ation}) for QED Hamiltonian are highly nonlinear so that the path integrals of the propagating function 
Eq.~(\ref{ppg}) cannot be carried out exactly. However, the similarity between Eq.~(\ref{qed_ation}) and Eq.~(\ref{fa_ation}) allows us to
find the nonperturbative renormalized electron Hamiltonian in non-rtelativistic QED.  

Note that the influence functional action Eq.~(\ref{qed_ation}) is a complex function. Its real part contains all the corrections to 
the electron Hamiltonian in non-rtelativistic  QED, which results in the renormalization of both the single electron energy and the electron-electron 
interaction. The imaginary part contains two decoherence sources. One contributes to the energy dissipation into the environment 
induced by the electron-photon interaction. The other contributes to the fluctuations arisen from the initial states of the thermal 
photonic reservoirs through the electron-photon interaction.
The influence functional action Eq.~(\ref{fa_ation}) shares the same property. Furthermore, it is not difficult to show that the last two 
terms in both Eqs.~(\ref{qed_ation}) and (\ref{fa_ation}) are pure imaginary so that they only contribute to the dissipation and 
fluctuation dynamics of the electron system.  The first two terms in both Eqs.~(\ref{qed_ation}) and (\ref{fa_ation}) can combine with the 
forward and backward bare system actions in Eq.~(\ref{qed_saction}) and (\ref{saction}), respectively, from which we can 
systematically find the general nonperturbative renormalized Hamiltonian of the system. 

Explicitly, let us first examine the generalized Fano-Anderson systems Eq.~(\ref{qth}) in Sec.~III. The renormalized 
system Hamiltonian can also be determined by the bare system Hamiltonian function in Eq.~(\ref{saction}) plus the 
first term in the influence functional action Eq.~(\ref{fa_ation}), i.e.,
\begin{align}
{\cal H}^r & [ \bm \alpha^\dag, \bm{\alpha}]={\cal H}[\bm \alpha^\dag, \bm{\alpha}] + \delta {\cal H}[\bm \alpha^\dag, \bm{\alpha}] \notag \\
&= \! \sum_i \!\varepsilon_i(\tau) \alpha^*_i(\tau) \alpha_i(\tau) \!-\! i\hbar \! \sum_{ij} \!\! \int_{t_0}^{\tau} \!\!\!\! d\tau' \! 
\alpha^*_i(\tau) g_{ij}(\tau,\tau') \alpha_j(\tau'). \label{fa_rsH1}
\end{align}
Note that the evolution of $\alpha_j(\tau)$ along the forward path is determined by \cite{Jin2010,Lei2012,Yang2015,Zhang2018} 
\begin{align}
\alpha_j(\tau)= \bm u_{jj'}(\tau,t_0)\alpha_j(t_0) + f_j(\tau)
\end{align}
where $\bm u_{jj'}(\tau,t_0)$ is the propagating Green function which obeys the integro-differential Dyson equation Eq.~(\ref{ute}). While, 
$f_j(\tau')$ is the noise source associated with the correlation Green function $\bm v(\tau,t)$ of Eq.~(\ref{vt}) that has no contribution
to the system Hamiltonian renormalization \cite{Yang2015,Zhang2018}. Thus, we can only take the part of the evolution 
$\alpha_j(\tau')$ that has the contribution to system Hamiltonian renormalization, i.e. 
\begin{align}
\alpha_j(\tau') & \propto \bm u_{jj'}(\tau',t_0)\alpha_j(t_0)  \notag \\
& \propto [\bm u(\tau',t_0)\bm u^{-1}(\tau,t_0)]_{jj'}\alpha_j(\tau).
\end{align}
The second line in the above expression also shows how the memory effect is taken into account. 
Using this result, Eq.~(\ref{fa_rsH1}) can be rewritten by
\begin{align}
{\cal H}^r[ & \bm \alpha^\dag, \bm{\alpha}]= \! \sum_{ij} \alpha^*_i(\tau)\varepsilon^r_{ij} (\tau,t_0) \alpha_{ij}(\tau) 
\end{align}
and
\begin{align}
\varepsilon^r_{ij} (\tau,t_0)& = \varepsilon_i(\tau) \delta_{ij} \!+\! \hbar {\rm Im} \!\! \int_{t_0}^{\tau} \!\!\!\!\! d\tau' 
 [\bm g(\tau,\tau')\bm u(\tau',t_0) \bm u^{-1}(\tau,t_0)]_{ij} \notag \\
 &= -\hbar {\rm Im}\big[\dot{\bm u}(\tau,t_0)\bm u^{-1}(\tau,t_0)\big]_{ij}
\end{align}
This is the same solution for the renormalized system Hamiltonian given by Eqs.~(\ref{fa_rH}) and (\ref{fa_re}) that we obtained 
after completely solved the propagating function Eq.~(\ref{rdmcr}) and derived the exact master equation Eq.~(\ref{EME}).

Thus, we can find the renormalized electron Hamiltonian in non-relativistic QED from the electron influence functional action  
Eq.~(\ref{qed_ation}) in the same way. The result is 
\begin{align}
{\cal H}^r( & \bm \xi^\dag(\tau), \bm{\xi}(\tau)) = \sum_{\bm p} \varepsilon_{\bm p} \xi^*_{\bm p}(\tau) \xi_{\bm p}(\tau)  \notag \\
& +\!\! \sum_{\bm p, \bm p', \bm q} \!\! U(\bm q) \xi^*_{\bm p+\bm q}(\tau) \xi^*_{\bm p'\!-\!\bm q}(\tau) \xi_{\bm p'}(\tau) 
\xi_{\bm p}(\tau) \notag \\
& \!- \! i\hbar \! \sum_{\bm{p}\bm{p}'\bm{k}} \! \int_{t_0}^{\tau} \!\!\! d\tau' \!
\xi^*_{\bm p}(\tau) \xi_{\bm p-\bm k}(\tau) G_{\bm k} (\tau,\tau') \xi^*_{\bm p'-\bm k}(\tau') \xi_{\bm p'}(\tau').  \label{reifor}
\end{align}
In Eq.~(\ref{reifor}), the first two terms are the bare electron Hamiltonian in Eq.~(\ref{qed_saction}), and the last term comes from the
 the first term in the electron influence functional action Eq.~(\ref{qed_ation}), as the renormalization effect arisen from the electron-photon interaction 
 after integrated out all the photonic modes.  Moreover, we can similarly introduce the two-electron propagating Green function 
 $\bm W_{\bm p, \bm p', \bm k}(t, t_0) \equiv \langle [c^\dag_{\bm p-\bm k}(t) c_{\bm p}(t),c^\dag_{\bm p'}(t_0) c_{\bm p'-\bm k}(t_0) ]\rangle $.
Consequently, we have 
 \begin{align}
 \xi^*_{\bm p-\bm k}(t) \xi_{\bm p}(t) \propto \sum_{\bm p''} \bm W_{\bm p, \bm p', \bm k} (t, t_0) 
 \xi^*_{\bm p'-\bm k}(t_0) \xi_{\bm p'}(t_0).
 \end{align}
 Then the renormalized electron Hamiltonian can be obtained as
 \begin{subequations}
 \label{qed_ren}
 \begin{align}
 H^r_{\rm electron} (t,t_0) & = \sum_{\bm p} \varepsilon^r_{\bm p}(t,t_0) c^\dag_{\bm p} c_{\bm p}  \notag \\
& +\!\! \sum_{\bm p, \bm p', \bm q} \!\! U^r_{\bm p'} (\bm q, t, t_0) c^\dag_{\bm p+\bm q} c^\dag_{\bm p'\!-\!\bm q} c_{\bm p'} c_{\bm p} ,
\end{align}
where
\begin{align}
&\varepsilon^r_{\bm p}(t,t_0)=\varepsilon_{\bm p} + \sum_{\bm p'}\delta U_{\bm p'}(\bm q,t,t_0) ,  \label{csee} \\
&U^r_{\bm p'}(\bm q, t, t_0)= U(\bm q)+\delta U_{\bm p'}(\bm q,t,t_0),
\end{align}
and
\begin{align}
\delta U_{\bm p'}(\bm q, t, t_0) = \hbar {\rm Im} \! \sum_{\bm q' \bm q''} \! \int_{t_0}^{t} \!\!\! d\tau
 & G_{\bm q} (t,\tau) \bm W_{\bm q', \bm q'', \bm q}(\tau,t_0) \notag \\
 & \times \bm W^{-1}_{\bm q'', \bm p', \bm q}(t,t_0) .
 \end{align}
\end{subequations}
The correction to the single electron energy in Eq.~(\ref{csee}) comes from the operator normal ordering of the renormalized electron-electron 
interaction in the last term of Eq.~(\ref{reifor}).  By calculating the  two-electron propagating Green function $\bm W(t, t_0)$ 
from the total non-relativistic QED Hamiltonian Eq.~(\ref{hqed}), the renormalized electron Hamiltonian and the electron reduced density
matrix can be obtained. From the renormalized electron Hamiltonian Eq.~(\ref{qed_ren}) and the reduced density matrix given by 
Eq.~(\ref{crhos_t})-(\ref{qed_ation}), the renormalized quantum thermodynamics, Eq.~(\ref{snd}) to Eq.~(\ref{ggss}) formulated in 
Sec.~III, can be directly extended to complicated interacting open quantum system. Of course, in practical, the two-electron propagating Green function 
$\bm W(t, t_0)$ is very difficult to be calculated, not like the systems of Eq.~(\ref{qth}) discussed in Sec. III where
the  general solution of the single particle Green function $\bm u(t,t_0)$ has been solved analytically in our previous work \cite{Zhang2012}. 

Nevertheless, Eqs.~(\ref{crhos_t}) to (\ref{qed_ation}) provide a full nonequilibrium electron-electron interaction theory rigorously 
derived from the non-relativistic QED theory after we integrated out exactly all the photonic modes. It can describe various 
nonequilibrium physical phenomena in many-body electronic systems based on the non-relativistic QED theory, where all the 
renormalization effects arisen from electron-photon interaction has been taken into account in the reduced density matrix. 
In practical, the reduced density matrix of Eq.~(\ref{crhos_t}) with Eqs.~(\ref{ppg}) to (\ref{qed_ation}) are still hard to be solved exactly, 
because the contributions from all the photon modes has been included exactly and it goes far beyond the perturbation 
expansion one usually used in many-body systems \cite{Thouless1972} and in quantum field theory \cite{Peskin1995}.    
In particular, when the Coulomb interaction dominates the electron-electron interaction, the system become a strongly correlated 
electronic system. Then further nonperturbative approximations and numerical methods have to be introduced to find properly the 
renormalized Hamiltonian and the reduced density matrix of the open system for the strong coupling quantum thermodynamics. 

In fact, the same problem also exists in the equilibrium physics, namely one cannot solve all the equilibrium physical problems exactly 
even though the Gibbs state is well defined under the equilibrium hypothesis of statistical mechanics. The typical example 
is the strongly-correlated electron systems, such as Hubbard model or quantum Heisenberg spin model, which are the 
approximation of the above nonequilibrium electron-electron interaction QED theory. But so far one is still 
unable to solve them exactly \cite{Nagaosa2010}. 
Therefore, how to practically solve nonequilibrium quantum thermodynamics for arbitrary system-environment interactions 
remains to be a challenge problem.  The closed time-path Green functions technique with the loop expansion to quantum 
transport phenomena developed by one of us long time ago  \cite{Zhang1992} could be a possible method for solving 
nonperturbatively the nonequilibrium quantum thermodynamics of strong interacting many-body systems, and we leave this 
problem for further investigation.  

\section{Conclusion and perspective}
In conclusion, we formulate the renormalization theory of quantum thermodynamics and quantum statistical mechanics  
based on the exact dynamic evolution of quantum mechanics for both weak and strong coupling strengths. 
For a class of generally solvable open quantum systems described by the generalized Fano-Anderson Hamiltonians, we show that the 
exact steady state of open quantum systems coupled to reservoirs through the particle exchange processes is a generalized 
Gibbs state.  The renormalized system Hamiltonian and the reduced density matrix are obtained nonperturbatively when we
traced over exactly all the reservoir states through the coherent state path integrals \cite{Zhang1990}.
Using the renormalized system Hamiltonian and introducing the renormalized temperature,  the exact steady state of the reduced 
density matrix can be expressed as the standard Gibbs state. The corresponding steady-state particle distributions obey 
the Bose-Einstein and the Fermi-Dirac distributions for bosonic and fermionic systems, respectively. In the very weak coupling 
limit, the renormalized system Hamiltonian and the renormalized temperature are reduced to the original bare Hamiltonian of the system and 
the initial temperature of the reservoir if it couples to a single reservoir. Thus, classical thermodynamics and statistical mechanics 
emerge naturally from the dynamics of open quantum systems. Thermodynamic laws and statistical mechanics principle are thereby deduced  
 from the dynamical evolution of quantum dynamics. If open quantum systems contain dissipationless localized bound states, 
thermalization cannot be reached. This is the solution to the long-standing problem in
thermodynamics and statistical mechanics that one has been trying to solve from quantum mechanics for a century. 

On the other hand, the renormalization theory presented in this work is nonperturbative. 
The traditional renormalization theory in quantum field theory and in many-body physics are build on perturbation 
expansions with respect to the interaction Hamiltonian.  
As we have systematically shown, the system Hamiltonian renormalization and the reduced density matrix are  finally
expressed in terms of the nonequilibrium Green functions. We have nonperturbatively derived the equation of motion for these 
nonequilibrium Green functions and obtained the general nonperturbation solution. We can easily reproduce the traditional 
perturbation renormalization theory by expanding order by order our solution with respect to the system-reservoir interaction.
Furthermore,  this nonperturbative renormalization theory also corresponds to the one-step renormalization in the framework of
Wilson renormalization group framework. The renormalization group is build through subsequent integrations of physical degrees 
of freedom from the higher energy scale to lower energy scale. For open quantum systems, instead of integrating out the higher 
energy degrees of freedom,  the dynamics is fully determined by nonperturbatively integrating out all the reservoir degrees of freedom at once
but including all energy levels from the low energy scale to the high energy scale of the reservoirs. Therefore, the underlying physical picture of 
our renormalization roots on the different physical basis. If the open quantum system interacts hierarchically with many 
reservoirs, then hierarchically tracing over all the reservoirs' states would lead to a new renormalization group theory to open
quantum systems that count all influences of hierarchical reservoirs on the system. 

As a consequence of the renormalized Hamiltonian and renormalized temperature, we find that the system can
become colder or hotter, as the coupling increases.  For fermion systems, as the coupling
increases, the renormalized energy levels can be increased or decreased, depending on the
dot energy levels are greater than or less than the center energy of the Lorentz-type spectral
density, but the renormalized temperature is always increased (becomes hotter). For boson systems
with the Ohmic-type spectral density, both the renormalized frequency and the renormalized
temperature always decrease (becomes colder) as the coupling increases, while for a Lorentz-type
spectral density, the renormalized frequency and
temperature will simultaneously decrease or increase, which is quite different from fermion systems.
This reveals the very flexible controllability for energy and heat transfers
between systems and reservoirs, and provides potential applications in building quantum
heat engines in strong coupling quantum thermodynamics.

\acknowledgments

This work is supported by Ministry of Science and Technology of Taiwan, Republic of China under
Contract No. MOST-108-2112-M-006-009-MY3.  

WMZ proposed the idea and formulated the theory, WMH performed the numerical calculations, 
WMH and WMZ discussed and analyzed the results, WMZ interpreted the physics and wrote the manuscript.

\appendix
\section{The derivation of Eq.~(\ref{ess}) from Eq.~(\ref{tsdm})}
In this appendix, we shall derive rigorously the reduced density matrix from the Gibbs state of the total system 
at the steady state by trace over all reservoir states. 

The total system is initially in an decoupled state between the system and the reservoir, given by Eq.~(\ref{itdm}).
After a long-time nonequilibrium evolution, the total system approaches to the steady equilibrium state which is the Gibbs 
state with a final equilibrium temperature denoted by $\beta_f=1/k_BT_f$, i.e.~Eq.~(\ref{tsdm}):
\begin{align}
\rho_{\rm tot}(t\!\rightarrow\!\infty)=\frac{1}{Z_{\rm tot}}e^{-\beta_f H_{\rm tot}},  \label{tsdm1}
\end{align}
where $H_{\rm tot}$ is the total Hamiltonian of Eq.~(\ref{fH}), and $Z_{\rm tot}=\Tr_{\s+\e}[e^{-\beta_f H_{\rm tot}}]$ is the 
corresponding partition function of the total system. This is also a direct consequence of statistical mechanics, namely when the 
total system is in equilibrium, its state is given by the Bolzertmann distribution, i.e., Eq.~(\ref{tsdm1}).
Because $H_{\rm tot}$ is a bilinear operator in terms of the bosonic creation and annihilation operators, Eq.~(\ref{tsdm1}) becomes
a Gaussian function in the coherent-state representation. Explicitly, Eq.~(\ref{tsdm1}) can be expressed as \cite{Huang2020},
\begin{align}
\bra z,\bm z |\rho_{\rm tot} | z',\bm z' \ket 
\!=\!\f{1}{Z_{\rm tot}} \exp \bigg\{\!\!\begin{pmatrix} z^\+ &\! \bm z^\+ \end{pmatrix}
\!\!\begin{pmatrix}\Omega_{\s\s}&\bm{\Omega}_{\s\e}\\\bm{\Omega}_{\e\s}&\bm{\Omega}_{\e\e}\end{pmatrix}
\!\!\begin{pmatrix}z' \\ \bm z' \end{pmatrix} \!\!\bigg\},
\end{align}
where 
\begin{align}
\begin{pmatrix}\Omega_{\s\s}&\bm{\Omega}_{\s\e}\\\bm{\Omega}_{\e\s}&\bm{\Omega}_{\e\e}\end{pmatrix}
=\exp \! \bigg[\!-\!\beta_f \begin{pmatrix} \hbar\omega_s & \hbar\bm V \\ \hbar\bm V^\dag & \hbar\bm \omega \end{pmatrix} \! \bigg].
\end{align}
Here, we have used the combined bosonic coherent state of the system plus the reservoir 
$|z, \bm z \rangle=\exp(z a^\+ + \sum_k z_k b^\+_k)|0\rangle$. We also used boldface to denote matrices and vectors.
As an explicit example, the vector $\bm z \equiv (z_{k_1} , z_{k_2}, . . .)$ and $z_{k_i}$ is a complex variable for reservoir boson mode $k_i$.

Taking the trace over all the reservoir modes, we obtain the reduced density matrix in the coherent state representation,
\begin{align}
\bra z |\rho_\s (t\!\rightarrow\!\infty) |z' \ket=\!\!\int \!\!d\mu(\bm z)\bra z,\bm z |\rho_{\rm tot}(t\!\rightarrow\!\infty)|z',\bm z \ket , \label{tracedm}
\end{align}
where $d\mu(\bm z)  = \prod_k \frac{dz^*_kdz_k}{2\pi i} e^{-|z_k|^{2}} $.
Use the Gaussian integral 
to complete the integration of Eq.~(\ref{tracedm}), we have 
\begin{align}
\bra z |\rho_\s (t\!\rightarrow\!\infty) |z' \ket=\frac{1}{Z_\s} e^{z^*\Omega_\s z'}
\end{align}
where $Z_\s=[1-\Omega_\s]^{-1}$ and $\Omega_\s=\Omega_{\s\s}+\bm{\Omega}_{\s\e}[\bm{I}-\bm{\Omega}_{\e\e}]^{-1}\bm{\Omega}_{\e\s}$.
On the other hand, the average particle number in the steady state is $\overline{n}=\tr_\s[ a^\+a\rho_\s (t\!\rightarrow\!\infty)]=
\!\!\int \!\!d\mu(\bm z)\bra z |a^\+a\rho_\s (t\!\rightarrow\!\infty) | z \ket$. The result is 
\begin{align}
\overline{n}(t\!\rightarrow\!\infty)
  &=\Omega_\s[1-\Omega_\s]^{-1}
\end{align}
Thus, the reduced density matrix becomes 
\begin{align}
\bra z |\rho_\s(t\!\rightarrow\!\infty)|z'\ket=\f{1}{Z_\s}e^{z^*\f{\bar{n}(t\!\rightarrow\!\infty)}{1+\bar{n}(t\!\rightarrow\!\infty)}z'}.
\end{align}
Using the fact that $\bra z |e^{fa^\dag a}|z'\ket= e^{z^*e^f z'}$, the above reduced density matrix can be written as a operator,  
\begin{align}
\rho_\s (t\!\rightarrow\!\infty) =\f{\exp\{\ln[\f{\bar{n}(t\!\rightarrow\!\infty)}{1+\bar{n}(t\!\rightarrow\!\infty)}]a^\+a\}}{1+\bar{n}(t\!\rightarrow\!\infty)},
\end{align}
which is the solution of Eq.~(\ref{ess}). This solution is obtained directly from Eq.~(\ref{tsdm1}). As we have found, the exact solution of  
$\bar{n}(t\!\rightarrow\!\infty)$ does not agree with the Bose-Einstein distribution if one takes the steady-state temperature as the initial reservoir 
temperature $T_0$ at strong coupling, even though the energy correction arisen from the strong coupling with the reservoir is properly included.  
Therefore, $T_f \neq T_0$ and temperature renormalization is necessary for strong coupling quantum thermodynamics. 
Note that a similar proof claimed in \cite{Subas2012} without providing any details should be incorrect.

\section{Exact solution and the steady state of open quantum systems with Eq.~(\ref{qth})}
In this Appendix, we present the general solution of open quantum systems with Eq.~(\ref{qth}). Without loss of the generality, 
the initial state of reservoir $\alpha$ can be assumed to be in a thermal state at temperature $T_{\alpha 0}$,  and the system can be 
an arbitrary state,
\begin{align*}
\rho_{\rm tot}(t_0)= \rho_\s(t_0) \otimes \rho_\e(t_0), ~~~ \rho_\e(t_0)= \prod_{\alpha}\frac{1}{Z_\alpha}e^{-\beta_\alpha H^\alpha_{\e}} .
\end{align*}
Here $Z_\alpha= \prod_{k}(1 \mp e^{-\beta_{\alpha 0} (\varepsilon_{\alpha k}-\mu_{\alpha 0})})^\mp$ is the partition 
function of reservoir $\alpha$. Different reservoir $\alpha$ could have differential initial
chemical potential $\mu_{\alpha 0}$ and different initial temperature $\beta_{\alpha 0} =1/k_B T_{\alpha 0}$. The up and down signs 
$\mp$ correspond to the reservoir being bosonic and fermionic systems, respectively.

After the initial time $t_0$, both the system and all reservoirs will evolve into a fully non-equilibrium state. 
For an arbitrary initial state $\rho_\s(t_0)$ of the system, the reduced density matrix at later time $t$ is 
defined by $\rho_\s(t)={\rm Tr}_\e [\rho_{\rm tot}(t)]$, which can be solved from Eq.~(\ref{fsLvNe}) in general.
To find the exact solution, we take the coherent state representation \cite{Zhang1990} again. Then,
the reduced density matrix $\rho_\s\left(  t\right) $ of Eq.~(\ref{fsLvNe}) can be expressed as
$\rho_\s\!(\boldsymbol{\alpha}_{t}, \boldsymbol{\alpha}_{t}^{\prime} , t)    
= \langle \boldsymbol{\alpha}_t|\rho_\s(t) | \boldsymbol{\alpha}^{\prime}_t \rangle$. 
Here we have used  the unnormalized  coherent state defined as: 
$|\boldsymbol{\alpha}\rangle=\exp(\sum_i \alpha_i a^{\dagger}_i)|0\rangle,
~d\mu\left( \boldsymbol{\alpha}\right)  = \prod_i g_i d\alpha^*_id\alpha_i
e^{-\left\vert \alpha_i\right\vert ^{2}} $,
the vector $ \boldsymbol{\alpha}\equiv (\alpha_1 , \alpha_2, . . .)$ is one-column matrix, and $\alpha_i$ 
are complex variables for bosons and Grassmannian variables for fermions with $g_i=1/2\pi i$ 
and $1$ in the Haar measure, respectively. Thus, the reduced density matrix in the coherent state representation 
can be expressed as 
\begin{align}
\rho_\s\!(\boldsymbol{\alpha}_{t}, \boldsymbol{\alpha}_{t}^{\prime} , t)    
=  \!\! \int   \!\! d\mu  \left(  \boldsymbol{\alpha}_{0}\right)  d\mu\left(\boldsymbol{\alpha}_{0}^{\prime}\right) 
 \rho_\s\left( \boldsymbol{\alpha}_{0},\boldsymbol{\alpha}_{0}^{\prime},t_{0}\right) 
 \notag \\    \times 
  \mathcal{J}\!\! \left(  \boldsymbol{\alpha}_{t},\boldsymbol{\alpha}_{t}^{\prime}, t; \boldsymbol{\alpha}_0,
  \boldsymbol{\alpha}_{0}^{\prime},t_0\right) .
\label{rho_t}%
\end{align}

The propagating function $\mathcal{J}\!\!\left( \boldsymbol{\alpha}^\dag_{t},
\boldsymbol{\alpha}_{t}^{\prime}, t; \boldsymbol{\alpha}_{0},{\boldsymbol{\alpha}'_0}^{\dag}, t_0\right)$
in Eq.~(\ref{rho_t}) can be obtained analytically after integrated exactly over all the environmental degrees 
of freedom using the coherent state path integrals. 
The result is \cite{Tu2008,Jin2010,Lei2012}
\begin{align}
\mathcal{J}(\bm{\alpha}^\dag_{t}, \bm {\alpha}'_{t}, t; &\bm \alpha_0,{\bm \alpha'_0}^{\dag}, t_0)= \int \! \mathcal{D}[\bm \alpha; 
\bm \alpha']\exp\Big\{\frac{i}{\hbar} \big(S_{s}[\bm \alpha^\dag,\bm \alpha] \notag \\
&-S^*_{s}[\bm \alpha'^\dag, \bm \alpha']+S_{\rm IF} [\bm \alpha^\dag,\bm \alpha; \bm \alpha'^\dag, \bm \alpha'] \big) \! \Big\} ,
\label{ppg1}
\end{align}
where $\mathcal{D}[\bm \alpha; \bm \alpha']$ is the path integral measure over the parameter space of the system coherent states 
$|\bm{\alpha}\rangle$. The bare system action in the coherent state basis is given by
\begin{align}
S_{s}[\bm \alpha^\dag,\bm \alpha] = & -{i\hbar \over 2}\big[\bm \alpha^\dag_t \bm{\alpha}(t) + \bm \alpha^\dag(t_0) \bm{\alpha}_{t_0} \big]  \notag \\
&+\!\! \int_{t_0}^{t}\!\! d\tau \Big\{ {i\hbar \over 2}
\big[\dot{\bm \alpha}^\dag(\tau) \bm{\alpha}(\tau)  -\bm \alpha^\dag(\tau) \dot{\bm{\alpha}}(\tau)\big] \notag \\
&~~~~~~~~~~~~~~~~~~~~ - {\cal H}(\bm \alpha^\dag(\tau), \bm{\alpha}(\tau)) \Big\}  \label{saction}
\end{align}
with the Hamiltonian function ${\cal H}(\bm \alpha^\dag(\tau), \bm \alpha(\tau))= \sum_i \varepsilon_i(\tau) \alpha^*_i(\tau) \alpha_i(\tau)$. 
The actions $S_{s}[\bm \alpha^\dag,\bm \alpha]$ and $S^*_{s}[\bm \alpha'^\dag,\bm \alpha']$ in Eq.~(\ref{ppg1} describe the forward and backward evolution of the 
system. The influence functional action $S_{\rm IF} [\bm \alpha^\dag,\bm \alpha; \bm \alpha'^\dag, \bm \alpha']$ represents all the influences of the 
reservoirs on the system after integrated out exactly all the environmental degrees of freedom. This procedure is called as the influence 
functional approach, proposed originally by Feynman and Vernon \cite{Feynman1963}. 
We extended the Feynman-Vernon's influence functional in terms of the coherent state path integrals so that the influence functional theory 
can be applied to both bosonic and fermionic environments \cite{Tu2008,Jin2010,Lei2012}. The resulting  influence functional action for  
the open quantum system Eq.~(\ref{qth}) is
\begin{widetext}
\begin{align}
S_{\rm IF}[\bm \alpha^\dag, \bm \alpha; \bm \alpha'^\dag, \bm \alpha']= 
\!\!  \int_{t_0}^{t}\!\!\! d\tau \bigg\{ i\hbar  \sum_{ij} & \Big[ \int_{t_0}^{\tau} \!\!\! d\tau' \!
\Big\{ \alpha^*_i(\tau) g_{ij}(\tau,\tau') \alpha_j(\tau') + \alpha'^*_i(\tau) g^*_{ij} (\tau,\tau') \alpha'_j(\tau') \notag \\
&\mp \!\! \int_{t_0}^{t} \!\!\! d\tau' \! \Big\{ \alpha'^*_i(\tau) g_{ij} (\tau,\tau') \alpha_j(\tau') \!-\! \big[\alpha^*_i(\tau) \mp \alpha'^*_i(\tau)\big]  
\widetilde{g}_{ij}(\tau,\tau') \big[\alpha_j(\tau') \mp \alpha'_j(\tau')\big] \!\Big\} \Big] \! \bigg\}, \label{fa_ation}
\end{align}
\end{widetext}
where the up and down signs of $\mp$ correspond respectively to the system being bosonic and fermionic. The system-reservoir 
correlation functions $g_{ij}(\tau,\tau')$ and $\widetilde{g}_{ij}(\tau,\tau')$ are given by Eq.~(\ref{giniti}). As one can see, 
both the bare system action and the action correction are quadratic with respect to the variables ${\alpha^*_i,\alpha_i}$
so that the path integrals of Eq.~(\ref{ppg1}) can be solved exactly. 
After a tedious calculation,
we obtain \cite{Tu2008,Jin2010,Lei2012}%
\begin{align}
\mathcal{J}( \boldsymbol{\alpha}^\dag_{t}, & \boldsymbol{\alpha}'_{t}, t;\boldsymbol{\alpha}_{0},
{\boldsymbol{\alpha}'_0}^{\dag}, t_0)  =  \big({\rm det}[\boldsymbol{w}(t)]\big)^{\pm 1}
 \notag \\ & \times 
\exp\Big\{ \boldsymbol{\alpha}_{t}^{\dag} \boldsymbol{J}_1(t,t_0)
\boldsymbol{\alpha}_{0}+ {\boldsymbol{\alpha}'}_{0}^{\dag}\boldsymbol{J}^\dag_1(t,t_0)  
\boldsymbol{\alpha}'_{t}  
\notag \\   & ~~~~~~~~~~~
\pm {\boldsymbol{\alpha}_0'}^\dag \boldsymbol{J}_3(t,t_0) \boldsymbol{\alpha}_{0} \pm
\boldsymbol{\alpha}_{t}^{\dag}\boldsymbol{J}_2(t)
\boldsymbol{\alpha}'_{t}\Big\} ,  \label{rdmcr}
\end{align}
where $\boldsymbol{w}\left(  t\right)  =[\boldsymbol{1}\pm \boldsymbol{v}\left(  t,t\right) ]^{-1}$, 
$\boldsymbol{J}_1(t,t_0) =\boldsymbol{w}(t)\boldsymbol{u}(t,t_0) $, $\boldsymbol{J}_2(t) = \boldsymbol{1}-\boldsymbol{w}(t)$ and 
$\boldsymbol{J}_3(t,t_0)=\boldsymbol{1}-\boldsymbol{u}^\dag(t,t_0)\boldsymbol{w}(t)\boldsymbol{u}(t,t_0) $, and
the up and down signs of $\pm$ correspond to the system being bosonic or fermionic systems again. 
The functions  $\boldsymbol{u}(  t,t_{0}) $ and $\boldsymbol{v}( t,t)  $ are the nonequilibrium dissipative 
particle-propagating Green's function and fluctuated particle-correlation Green's function, respectively. They 
are determined by Eq.~(\ref{uvgf}).  Equations (\ref{rho_t}) and (\ref{rdmcr}) give the exact solution of the reduced 
density matrix in coherent state representation for the open quantum systems by Eq.~(\ref{qth}).
From Eqs.~(\ref{rho_t})-(\ref{rdmcr}), we have derived rigorously the exact master equation Eq.~(\ref{EME}) 
\cite{Tu2008,Jin2010,Lei2012}.

Based on the general solution of the nonequilibrium Green functions we obtained recently \cite{Zhang2012}, 
if there is no localized bound states (modes), the dissipative propagating Green function 
vanishes in the steady-state limit, namely
\begin{align}
\boldsymbol{u}(t \rightarrow \infty,t_0) = 0 .
\end{align}
This solution is valid for arbitrary continuous spectral density matrix of multiple reservoirs $J_{\alpha ij}(\omega)$ 
that cover every point of the whole energy frequency domain.
Then the coefficients in  the propagating function of the reduced density matrix, Eq.~(\ref{rdmcr}), is largely simplified:
$\boldsymbol{J}_1(t,t_0) =0 $, $\boldsymbol{J}_2(t) = \bm{1}-\bm{w}(t) = \pm \bm{v}(t,t)/[\bm{1}\pm \bm{v}(t,t)]$ and 
$\bm{J}_3(t,t_0)=\bm{1}$. Thus, the propagating function is simply reduced to
\begin{align}
\mathcal{J}( \bm{\alpha}^\dag_{t},  \bm{\alpha}'_{t},&  t\rightarrow \infty; \bm{\alpha}_{0},
{\bm{\alpha}'_0}^{\dag}, t_0)  = \lim_{t\rightarrow\infty} \big({\rm det}[\bm{w}(t)]\big)^{\pm 1} 
\notag \\ & \times 
\exp\Big\{ \pm {\bm{\alpha}_0'}^\dag \bm{\alpha}_{0} \pm
\bm{\alpha}_{t}^{\dag}  [\bm{1}-\bm{w}(t)]
\bm{\alpha}'_{t}\Big\} .  \label{rdmcr1}
\end{align}
Substituting this result into Eq.~(\ref{rho_t}), we obtain the exact steady-state reduced density matrix,
\begin{align}
\rho_\s(\bm{\alpha}_{t}, & \bm{\alpha}_{t}^{\prime} , t \rightarrow \infty )    
= \! \lim_{t\rightarrow\infty} \! \int   \!\! d\mu  \left(  \bm{\alpha}_{0}\right)  d\mu\left(\bm{\alpha}_{0}^{\prime}\right) 
 \rho_\s\left( \bm{\alpha}_{0},\bm{\alpha}_{0}^{\prime},t_{0}\right) 
 \notag \\  & \times \!  
 \big({\rm det}[\bm{w}(t)]\big)^{\pm 1} \!\!
  \exp\Big\{ \pm {\bm{\alpha}_0'}^\dag \bm{\alpha}_{0} \pm
\bm{\alpha}_{t}^{\dag} [\bm{1}-\bm{w}(t)] \bm{\alpha}'_{t}\Big\}  .
\label{rho_infty}%
\end{align}
Notice the normalization condition
\begin{align}
\int   \!\! d\mu  \left(  \bm{\alpha}_{0}\right)  d\mu\left(\bm{\alpha}_{0}^{\prime}\right) 
 \rho_\s\left( \bm{\alpha}_{0},\bm{\alpha}_{0}^{\prime},t_{0}\right)  \exp\Big\{ \!\! 
 \pm {\bm{\alpha}_0'}^\dag \bm{\alpha}_{0} \Big\} =1,
\end{align}
we have 
\begin{align}
\langle \bm{\alpha}_{t}| & \rho_\s\left( t \rightarrow \infty \right)   | \bm{\alpha}'_{t} \rangle 
\notag \\& 
=\lim_{t\rightarrow\infty} \big({\rm det}[\bm{w}(t)]\big)^{\pm 1} \!\!  \exp\Big\{ \pm
\bm{\alpha}_{t}^{\dag}  [\bm{1}-\bm{w}(t)] \bm{\alpha}'_{t}\Big\}   
\notag \\&
= \lim_{t\rightarrow\infty} \Big({\rm det}[ \frac{\bm{1}}{\bm{1} \pm \bm{v}(t,t)}]\Big)^{\pm 1} \!\!  \langle \bm{\alpha}_{t}|
 \frac{\bm{v}(t,t)}{\bm{1} \pm \bm{v}(t,t)}  \bm{\alpha}'_{t} \rangle.  \label{csrdm}
\end{align}
This shows that the steady-state reduced density matrix is independent of its initial states, 
as a consequence of thermalization. Equation (\ref{csrdm}) directly results in the operator form of the 
steady-state reduced density matrix
\begin{align}
\rho_\s\!\left( t \rightarrow \infty \right)= & \lim_{t\rightarrow\infty} \Big(\frac{1}{\det[\bm{1} \pm \bm{v}(t,t)]}\Big)^{\pm 1}
\notag \\ & \times \!\exp \Big\{  \bm{a}^\dag \!\Big(\! \ln  \frac{\bm{v}(t,t)}{\bm{1} \pm \bm{v}(t,t)} \Big)\bm{a} \Big\},
\end{align}
where $\bm a \equiv (a_1,a_2, \cdots, a_N)^T$ is a one-column matrix operator. 
This is the exact steady-state solution of Eq.~(\ref{gss}), which is also recently derived from the
general solution of the reduced density matrix \cite{Xiong2020}.  As one can see, the solution of Eq.~(\ref{ess}) 
derived alternatively in Appendix A is a special case of the above general solution.
We should also point out that the above solution remains the same for the initial coupled 
system-reservoir state \cite{Yang2015,Huang2020}.


\begin{thebibliography}{99}
\bibitem{Coulomb_gauge} Usually one takes the Coulomb gauge $\nabla \cdot \bm A=0$, then the Hamiltonian should 
contain a Coulomb interaction between electrons: $U\!\!=\!\! \frac{e^2}{4\pi \epsilon_0}\! \! \int 
\!\! d^3r d^3r' \frac{\varrho(\bm r)\varrho(\bm r')}{|\bm r-\bm r'|}$,  where $\varrho(\bm r)$ is the electron 
charge density at position $\bm r$. This is the origin of the second term in Eq.~(\ref{hqed}).
\end{thebibliography}
\end{document}